\documentclass{aa}

\usepackage[varg]{txfonts}
\usepackage[dvips]{color}
\usepackage{graphicx,subfigure}
\usepackage{amsmath}
\usepackage{epsfig,amsmath,amssymb}

\begin{document}

%
\titlerunning{Interacting MHD superbubbles}
\authorrunning{Ntormousi, Dawson et al.}
\title{The role of magnetic fields in the structure and interaction of supershells}
\author{Evangelia Ntormousi\inst{1}, Joanne R. Dawson \inst{2,3}, Patrick Hennebelle \inst{1,4}, Katharina Fierlinger \inst{5,6}}
\date{Received -- / Accepted --}

\institute{
Laboratoire AIM, 
Paris-Saclay, CEA/IRFU/SAp - CNRS - Universit\'e Paris Diderot, 91191, 
Gif-sur-Yvette Cedex, France \\
\and
Department of Physics and Astronomy and MQ Research Centre in Astronomy, Astrophysics and Astrophotonics, Macquarie
University, NSW 2109, Australia \\
\and
Australia Telescope National Facility, CSIRO Astronomy and Space Science, PO Boc 76, Epping, NSW 1710, Australia \\
\and
LERMA (UMR CNRS 8112), Ecole Normale Sup\'erieure, 75231 Paris Cedex, France \\
\and
University Observatory Munich, Scheinerstr. 1, D-81679 M\"unchen, Germany \\
\and
Excellence Cluster Universe, Technische Universit\"at M\"unchen, Boltzmannstr. 2, D-85748, Garching, Germany \\
}

\abstract
{
Large-scale shocks formed by clustered feedback of young OB stars
are considered an important source of mechanical energy for the interstellar medium (ISM) 
and a trigger of molecular cloud formation. 
Their interaction sites are locations where kinetic energy and magnetic fields are redistributed between ISM phases.
}
{
In this work we address two questions, 
 both involving the role of galactic magnetic fields in the dynamics of supershells and their interactions.
On the one hand, we study the effect of the magnetic field on the expansion and 
fragmentation of supershells and, on the other hand, we look for the signatures of supershell collisions
on dense structures and on the kinetic and magnetic energy distribution of the ISM.  
}
{
We performed a series of high-resolution, three-dimensional simulations of colliding supershells.  
The shocks are created by time-dependent feedback and evolve in a diffuse turbulent environment that is either unmagnetized or has
different initial magnetic field configurations. 
}
{
In the hydrodynamical situation, the expansion law of the superbubbles is consistent with the radius-time relation 
$R\propto t^{3/5}$ that is theoretically predicted for wind-blown bubbles.  The supershells fragment over their entire surface into small dense
clumps that carry more than half of the total kinetic energy in the volume.  However, 
this is not the case when a magnetic field is introduced, either in the direction of the collision or perpendicular to the collision.
In both situations, the shell surfaces are more stable to dynamical instabilities. When the magnetic field opposes the collision, the expansion law
of the supershells also becomes significantly flatter than in the hydrodynamical case.
Although a two-phase medium arises in all cases, in the magnetohydrodynamical (MHD) simulations  
the cold phase is limited to lower densities and the cold clumps are located further away from the shocks with respect to the hydrodynamical simulations.
}
{
For the parameters we explored, self-gravity has no effect on either the superbubble expansion
or the shock fragmentation. 
In contrast, a magnetic field, whether mostly parallel or mostly perpendicular to the collision axis, 
causes a deceleration of the shocks, deforms them significantly, and largely suppresses the formation
of the dense gas on their surface.  
The result is a multi-phase medium in which the cold clumps are not spatially correlated with the supershells.  
}


\maketitle

\begin{figure*}[!th]
\begin{center}
    \includegraphics[width=0.47\linewidth]{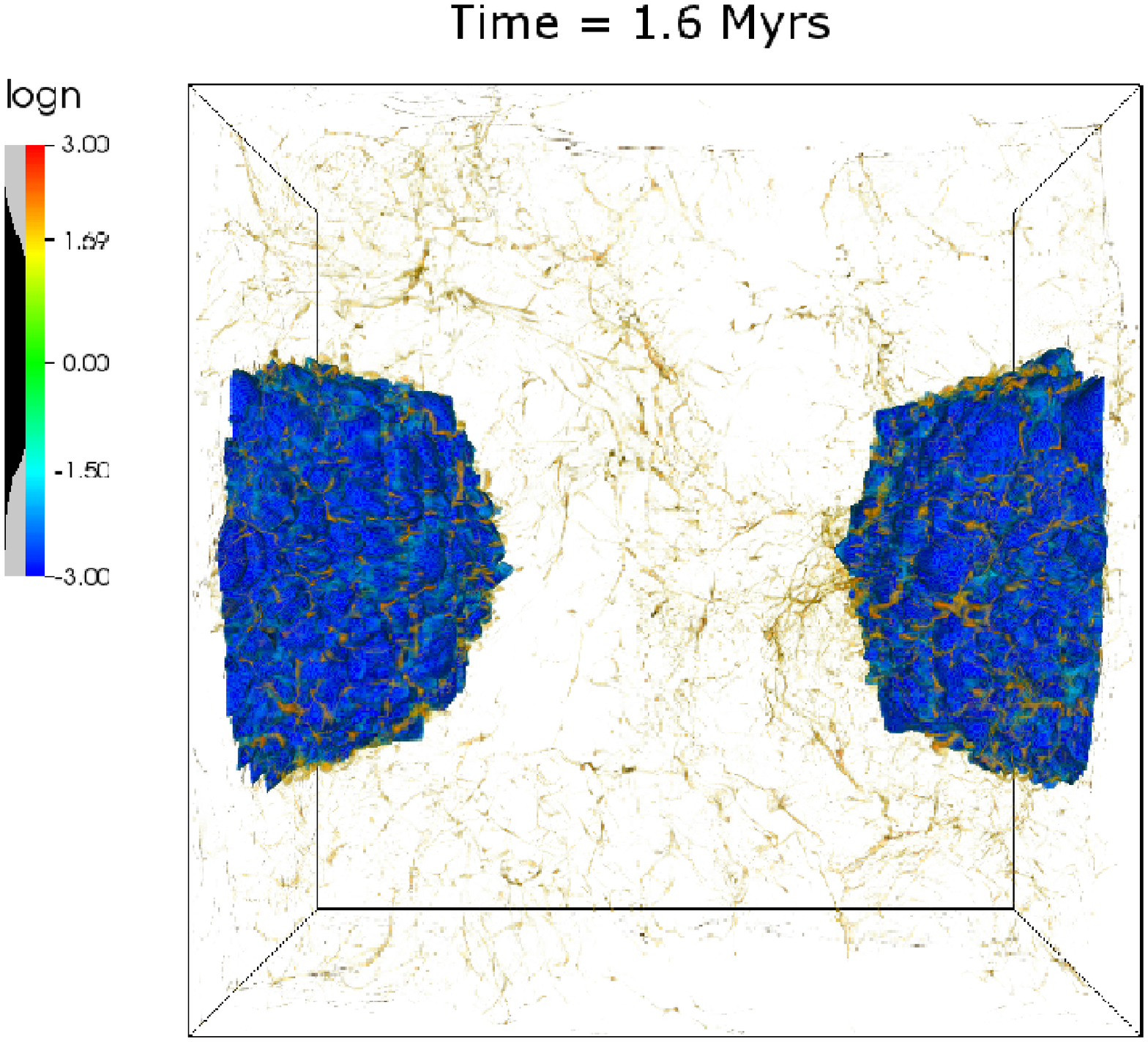}
    \includegraphics[width=0.47\linewidth]{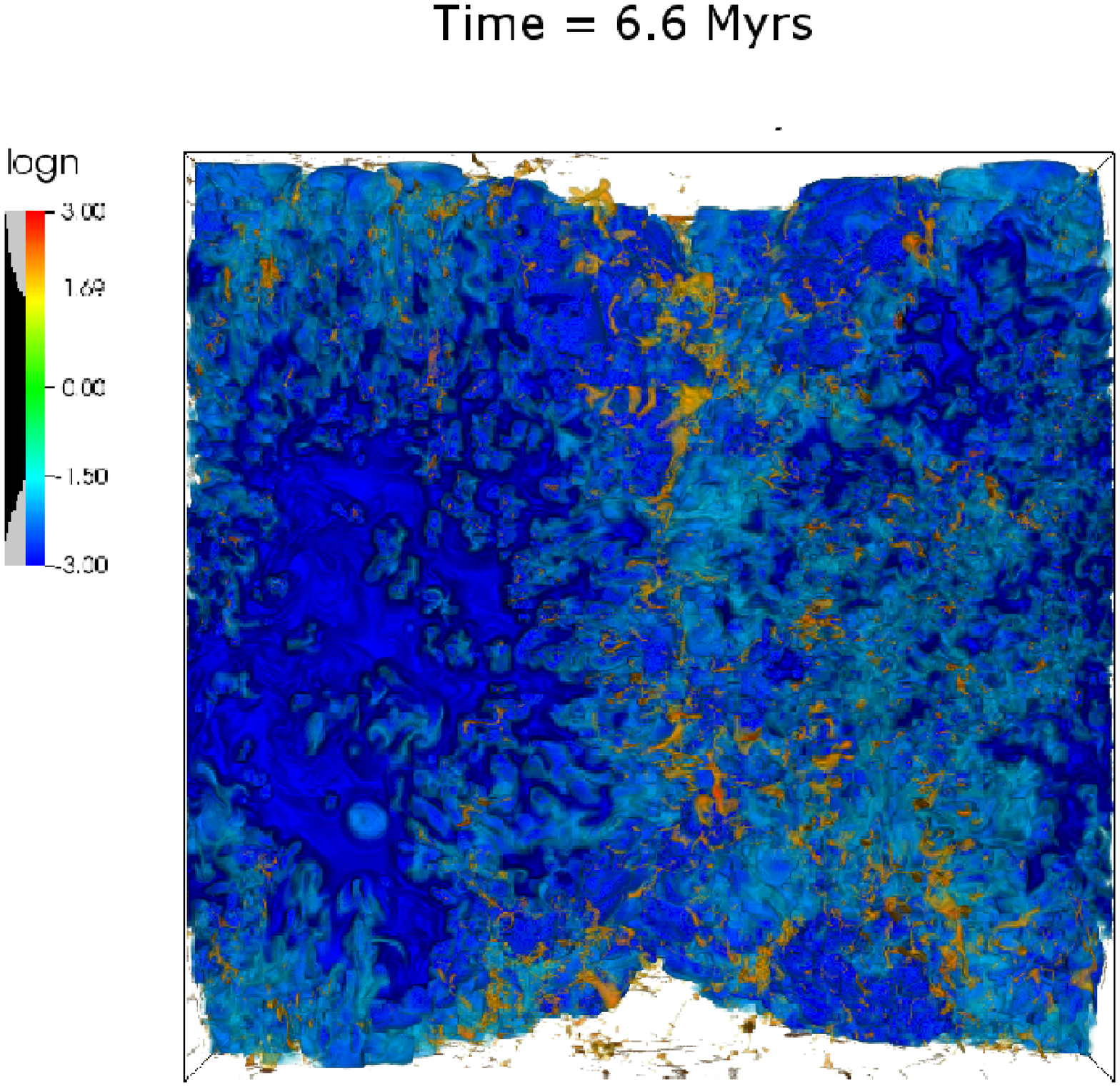}
    \includegraphics[width=0.45\linewidth]{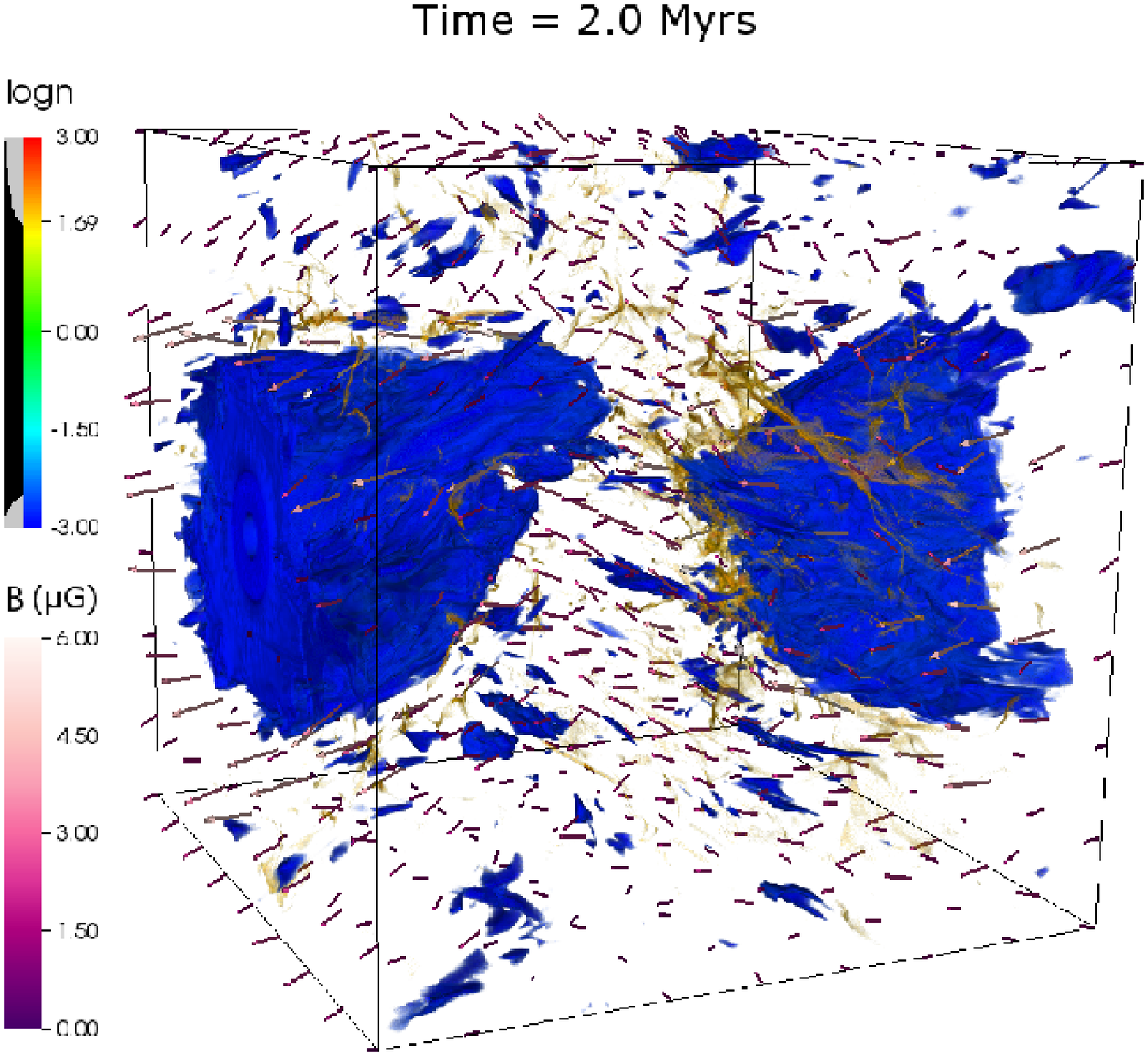}
    \includegraphics[width=0.45\linewidth]{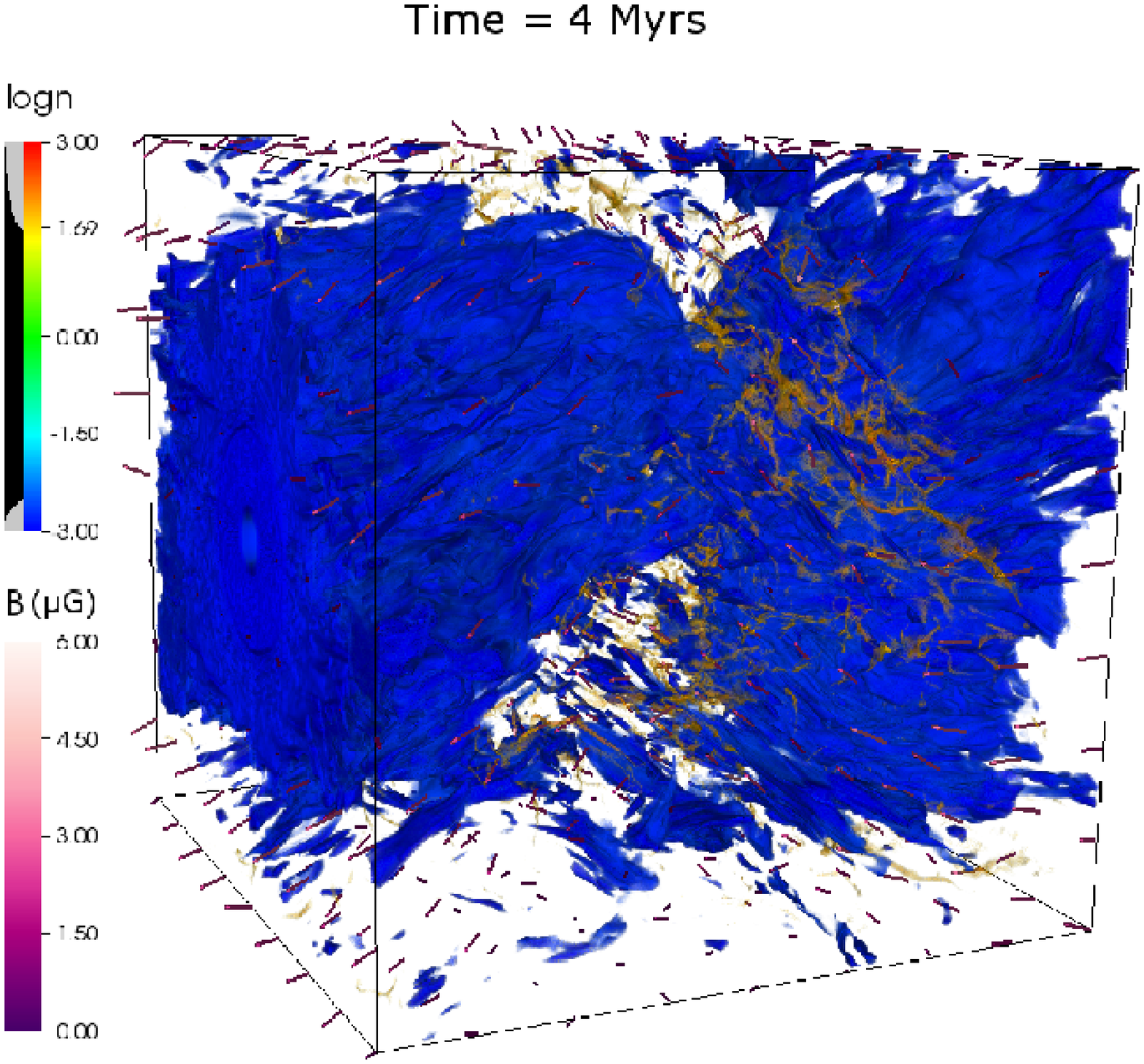}
    \includegraphics[width=0.45\linewidth]{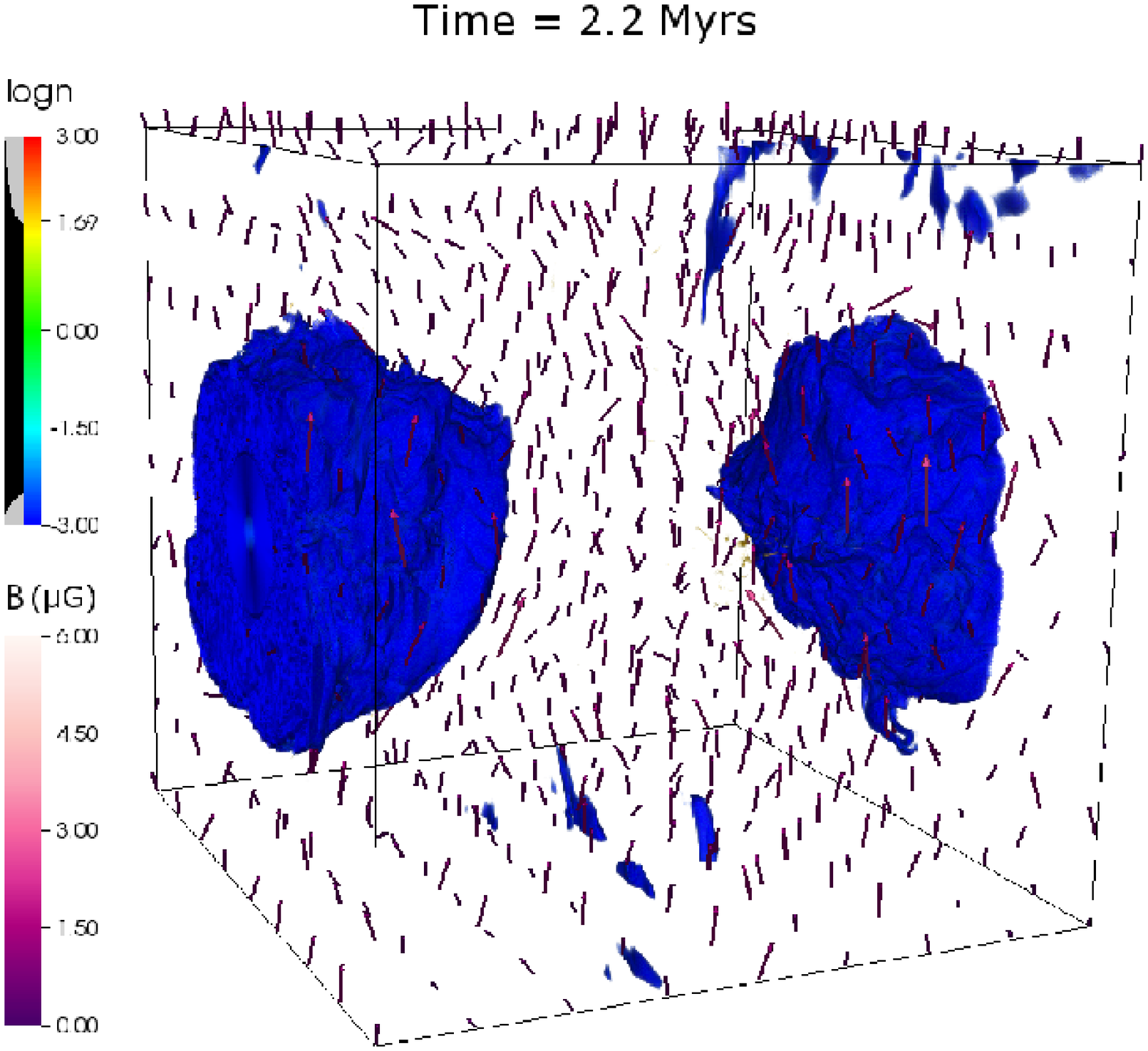}
    \includegraphics[width=0.45\linewidth]{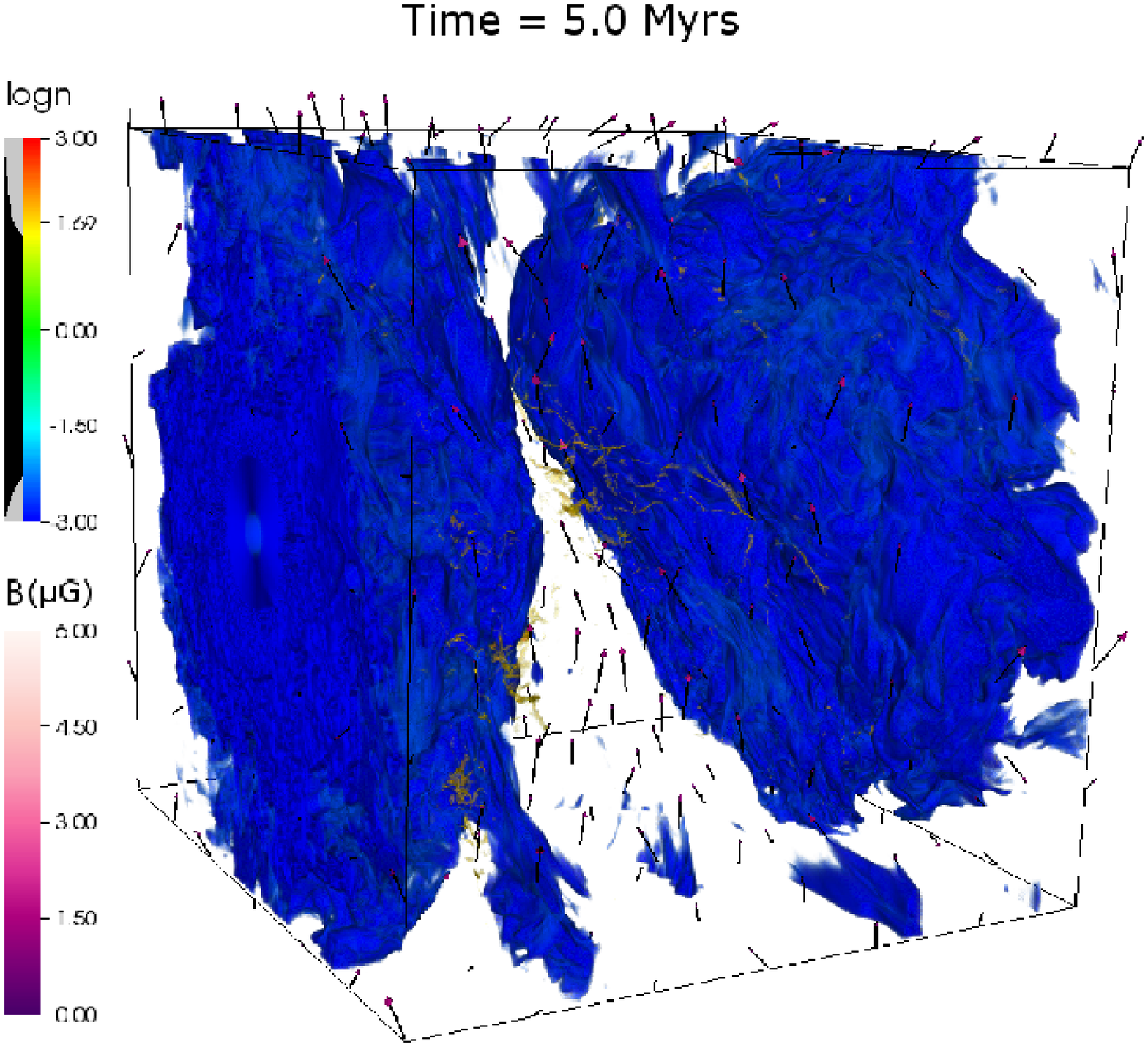}
    \caption{Volume rendering of the logarithm of the density field in the hydrodynamical run \textit{hyd} (top), and the two MHD runs, \textit{mhd1r} (middle), and \textit{mhd1t} (bottom).  Time increases from left to right, as indicated in the plot labels.
The warm medium was rendered transparent below 50 cm$^{-3}$ to allow a clear view of the dense clumps.  Transparency is indicated by the gray/black curve along the color bar.}  
     \label{volume_rendering_all}
\end{center}
\end{figure*}


\begin{figure*}[!th]
\begin{center}
    \includegraphics[width=0.44\linewidth]{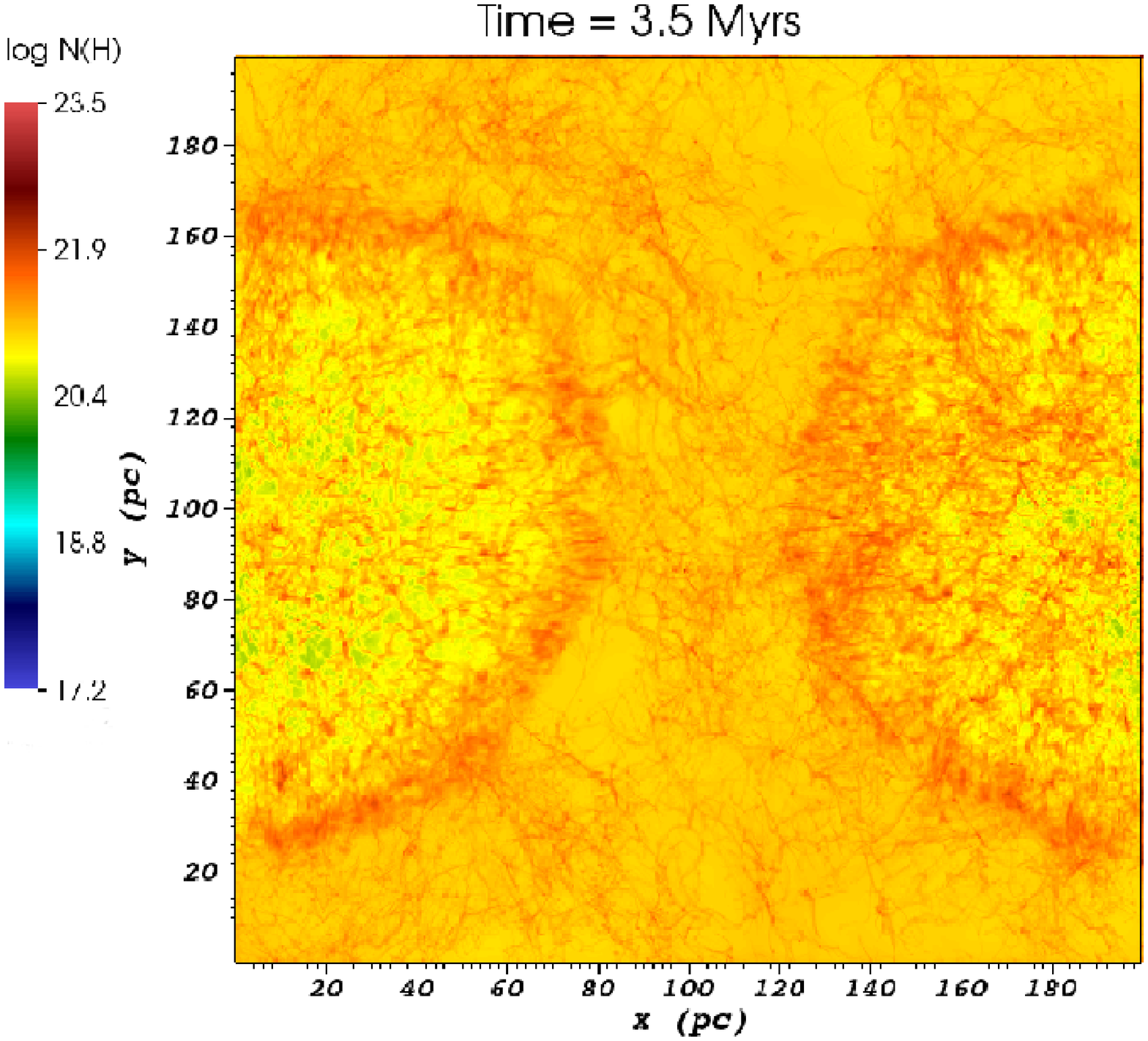}
    \includegraphics[width=0.44\linewidth]{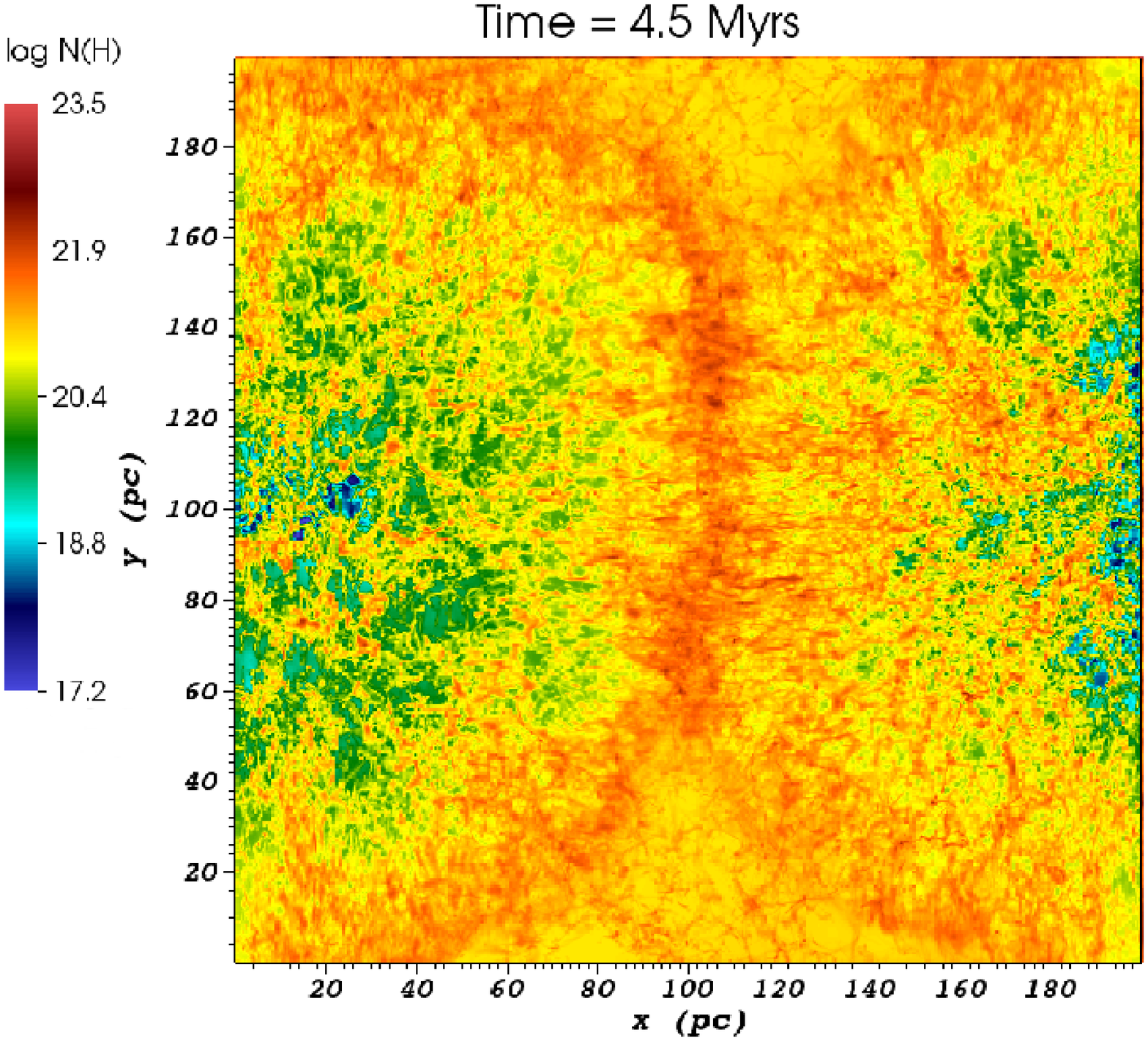}
   \includegraphics[width=0.44\linewidth]{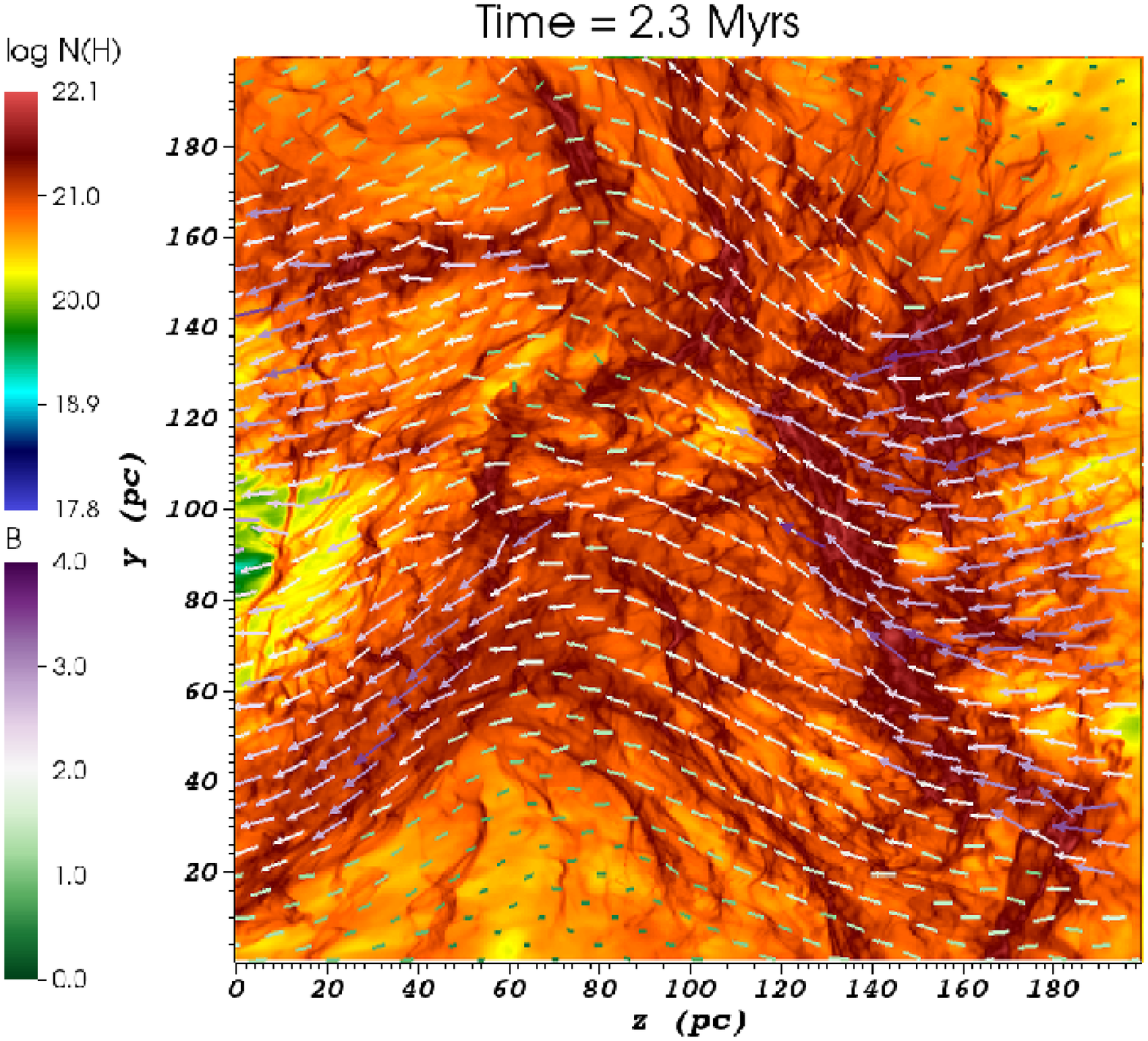}   
    \includegraphics[width=0.44\linewidth]{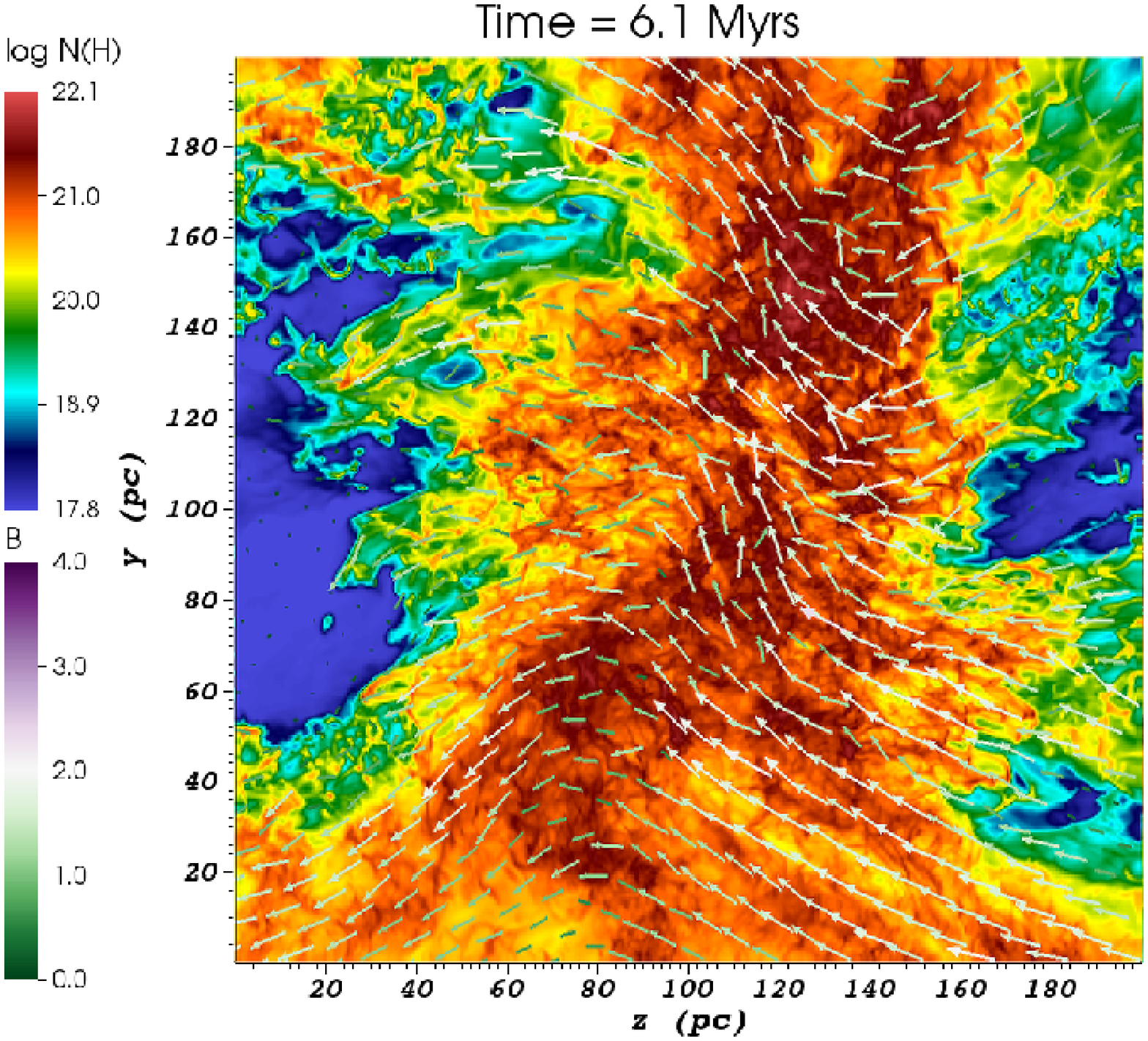}
    \includegraphics[width=0.44\linewidth]{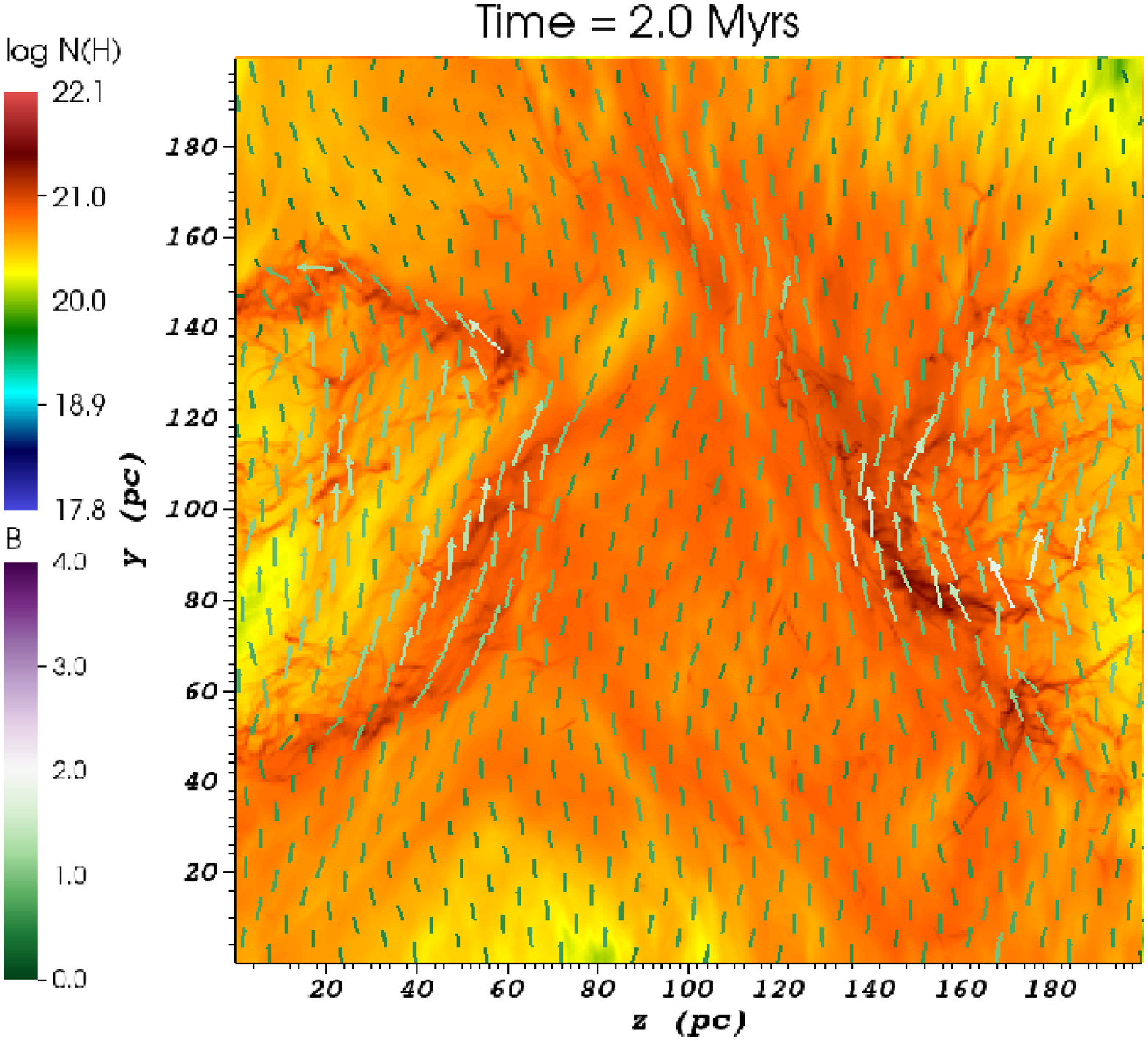}
   \includegraphics[width=0.44\linewidth]{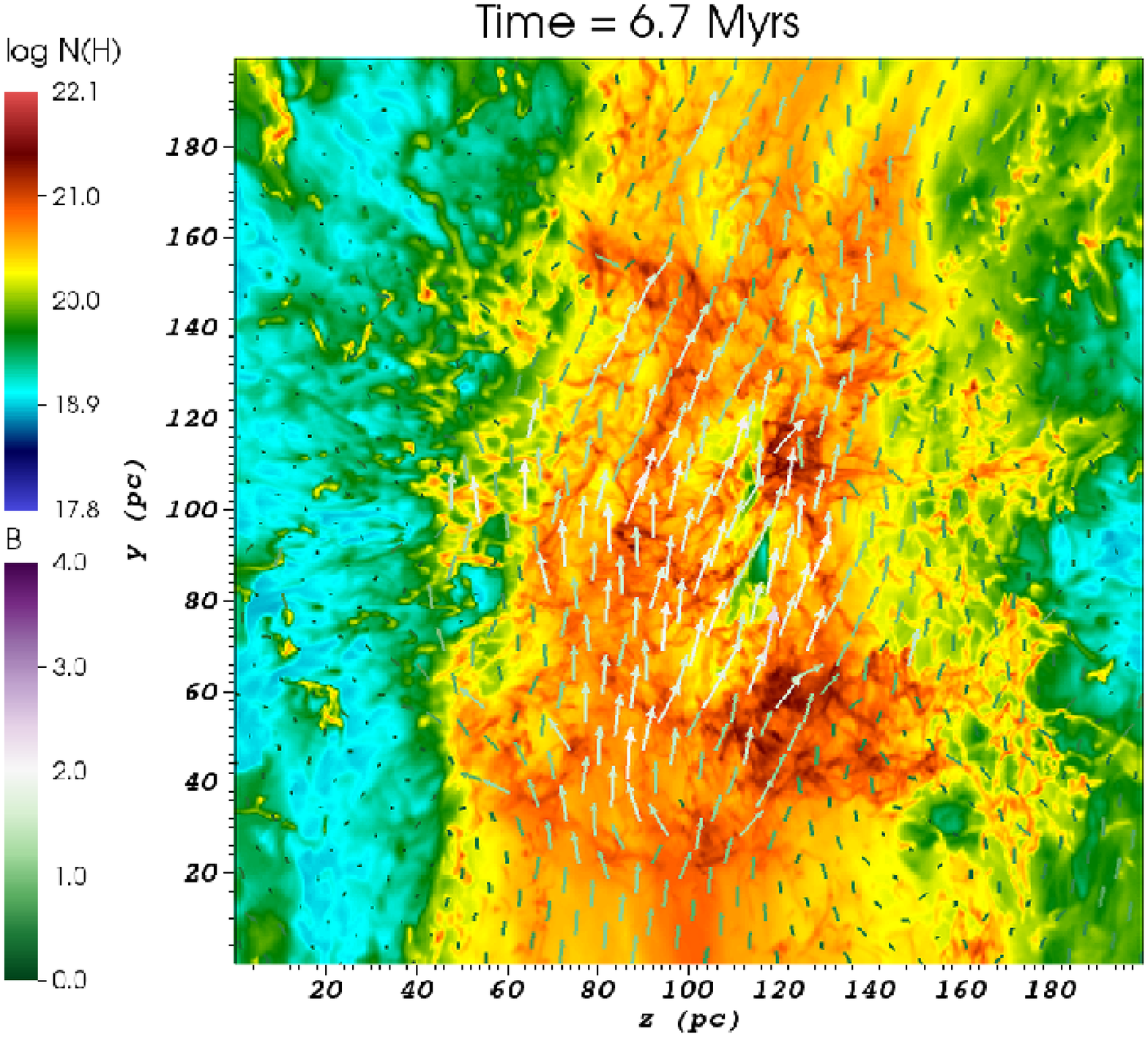}

    \caption{Column density contours for the pure hydrodynamic run \textit{hyd} (top) and the two MHD runs, \textit{mhd1r} (middle) and \textit{mhd1t} (bottom).  
In runs \textit{mhd1r} and \textit{mhd1t} the collision occurs along the z-axis.  The initial magnetic field of run \textit{mhd1r} is also along the z-axis, while that of \textit{mhd1t} is along the y-axis. Time increases from left to right, as indicated in the plot labels. The mass-weighted projected magnetic field is shown with arrows, color coded according to its magnitude in $\mu$G.}
     \label{shells_col_den_all}
\end{center}
\end{figure*}

\begin{figure*}[!th]
    \includegraphics[width=0.33\linewidth]{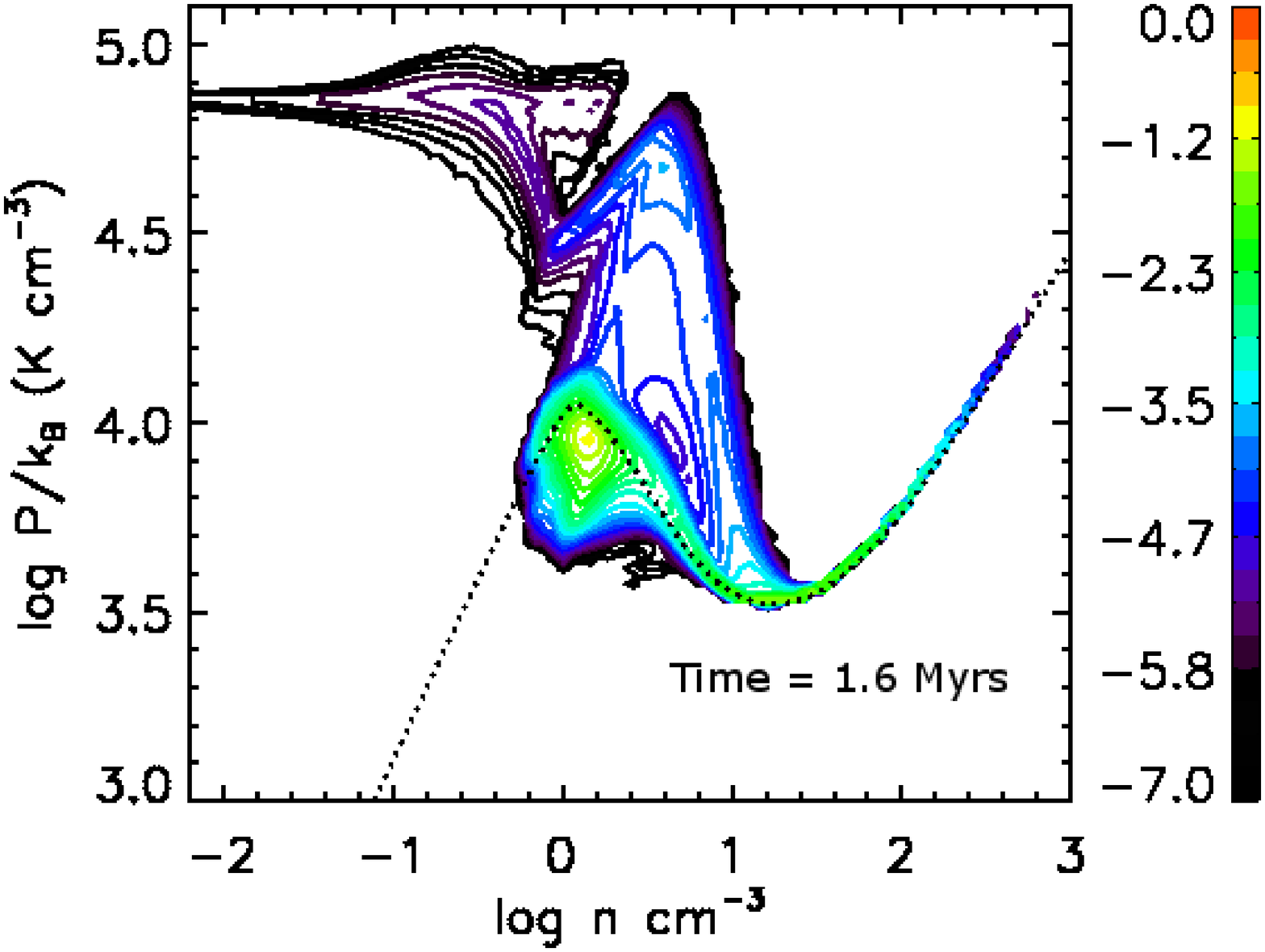}
   \includegraphics[width=0.33\linewidth]{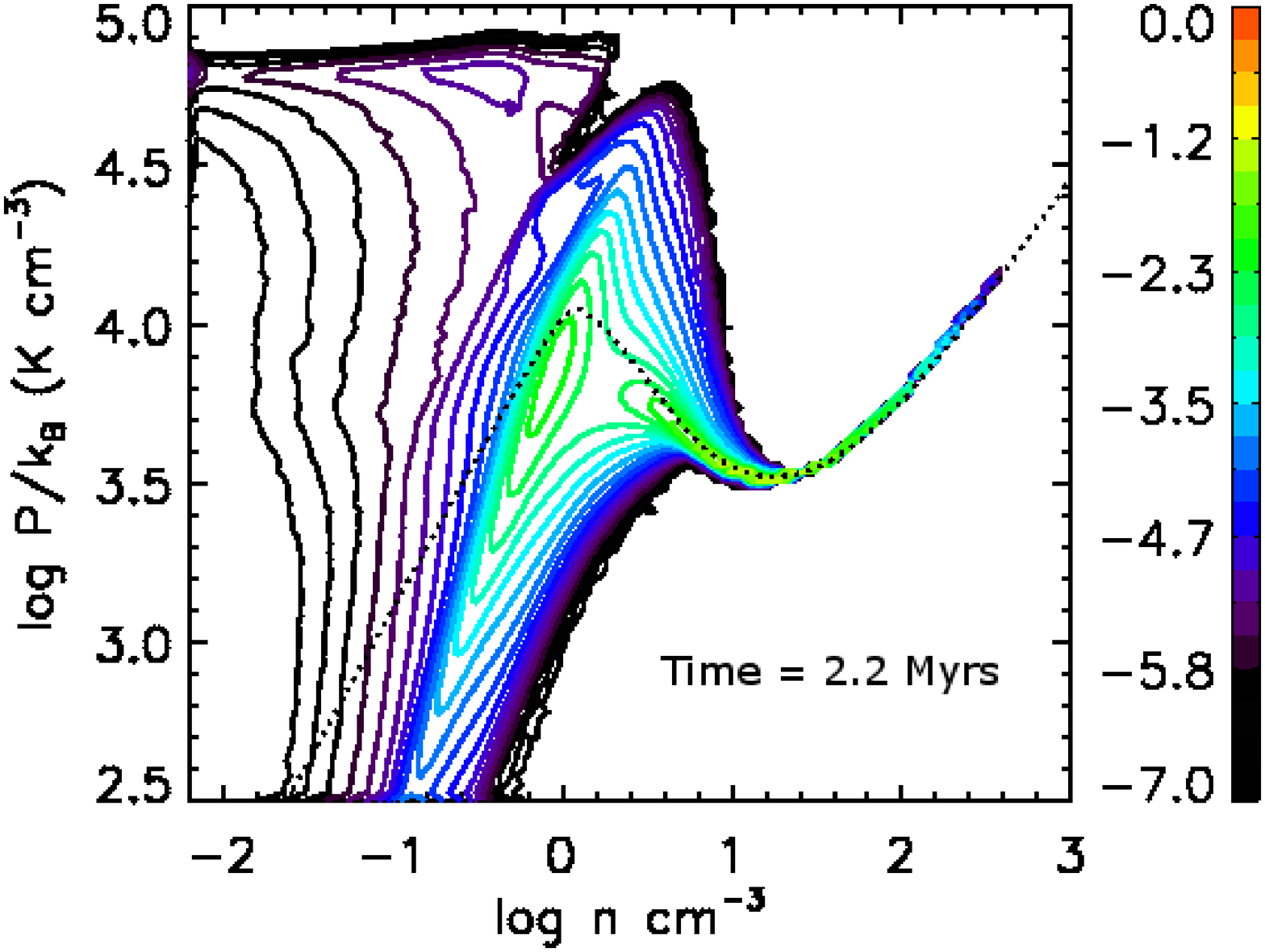}
   \includegraphics[width=0.33\linewidth]{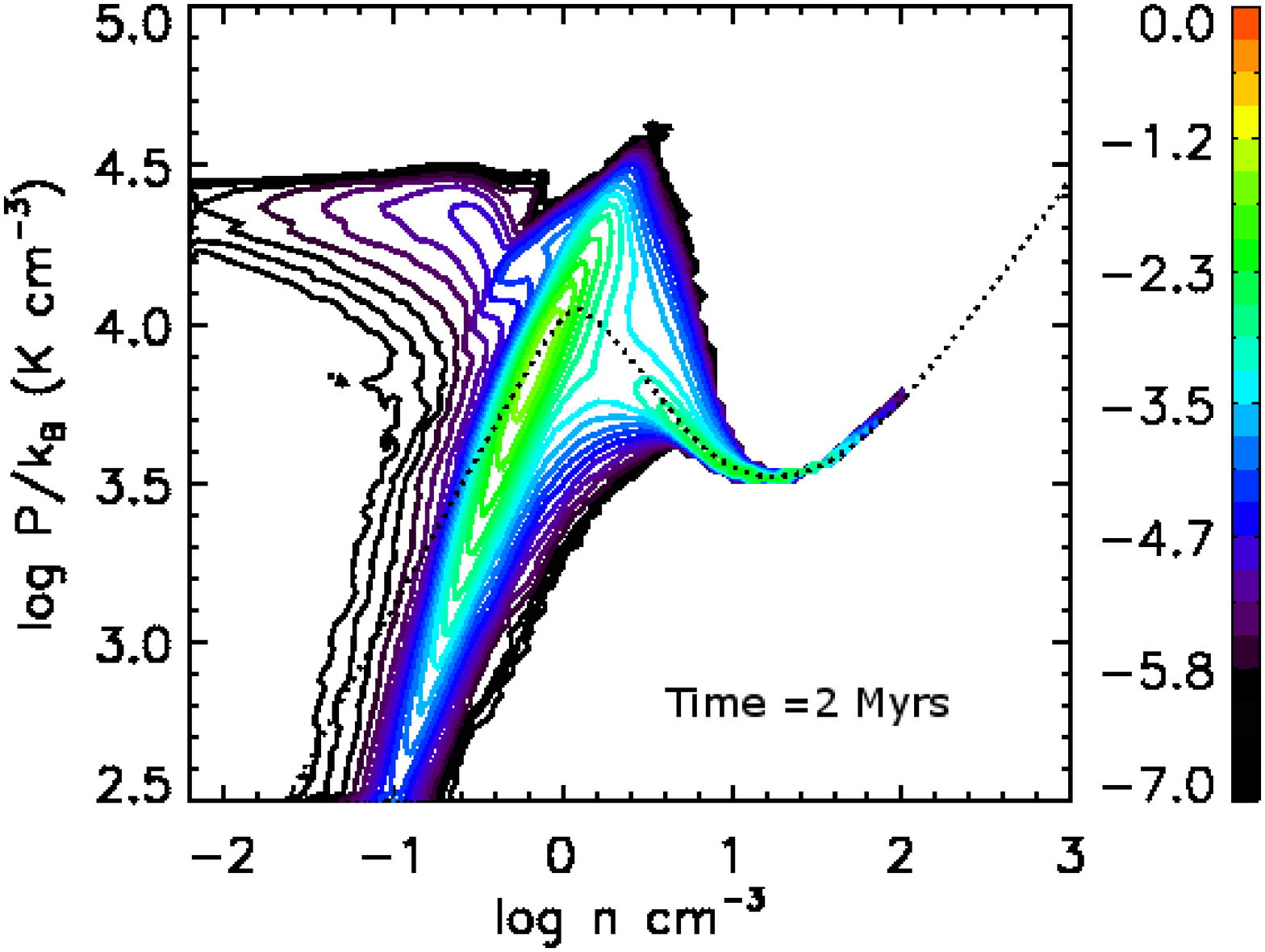}
    \includegraphics[width=0.33\linewidth]{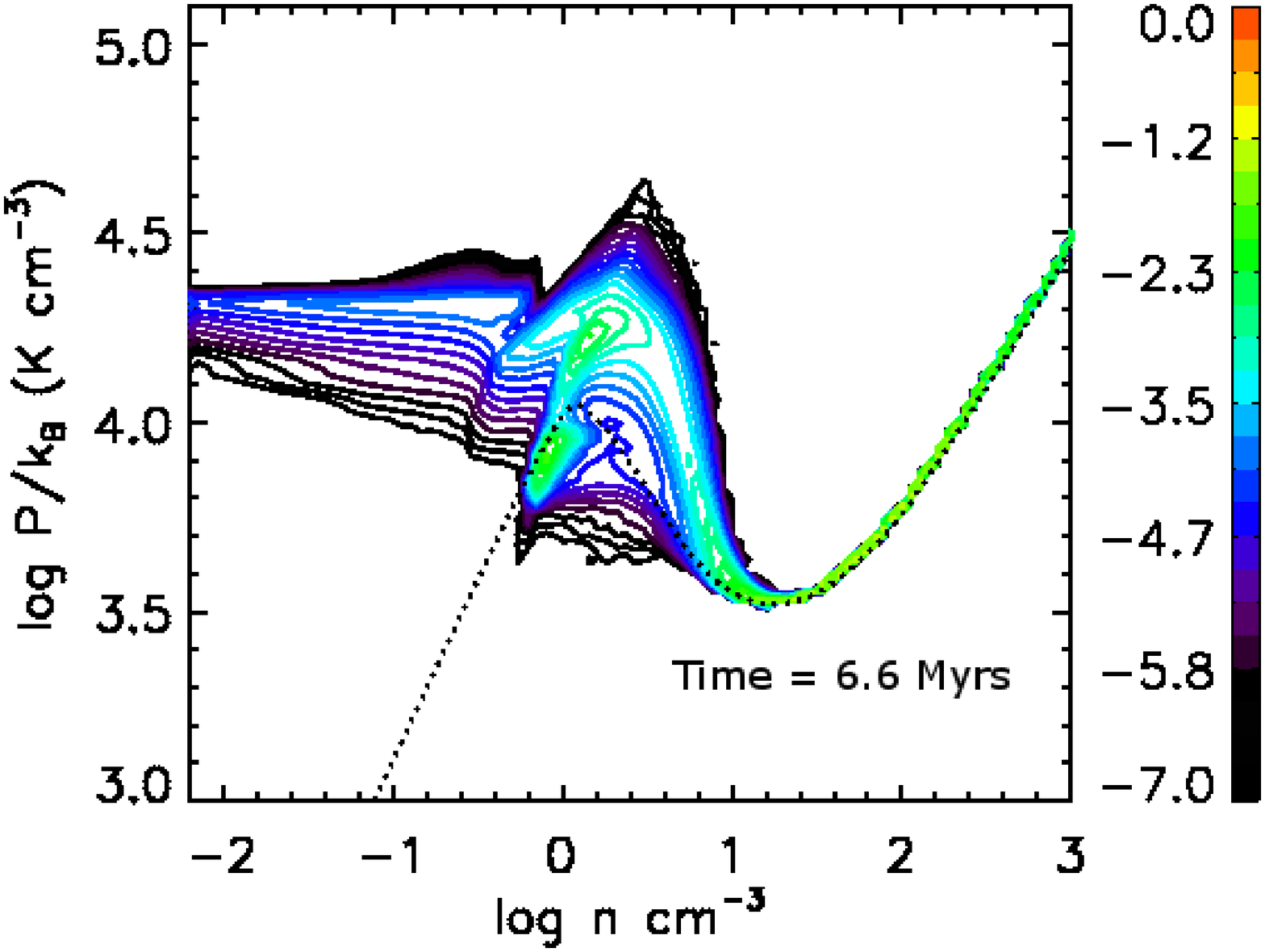}
    \includegraphics[width=0.33\linewidth]{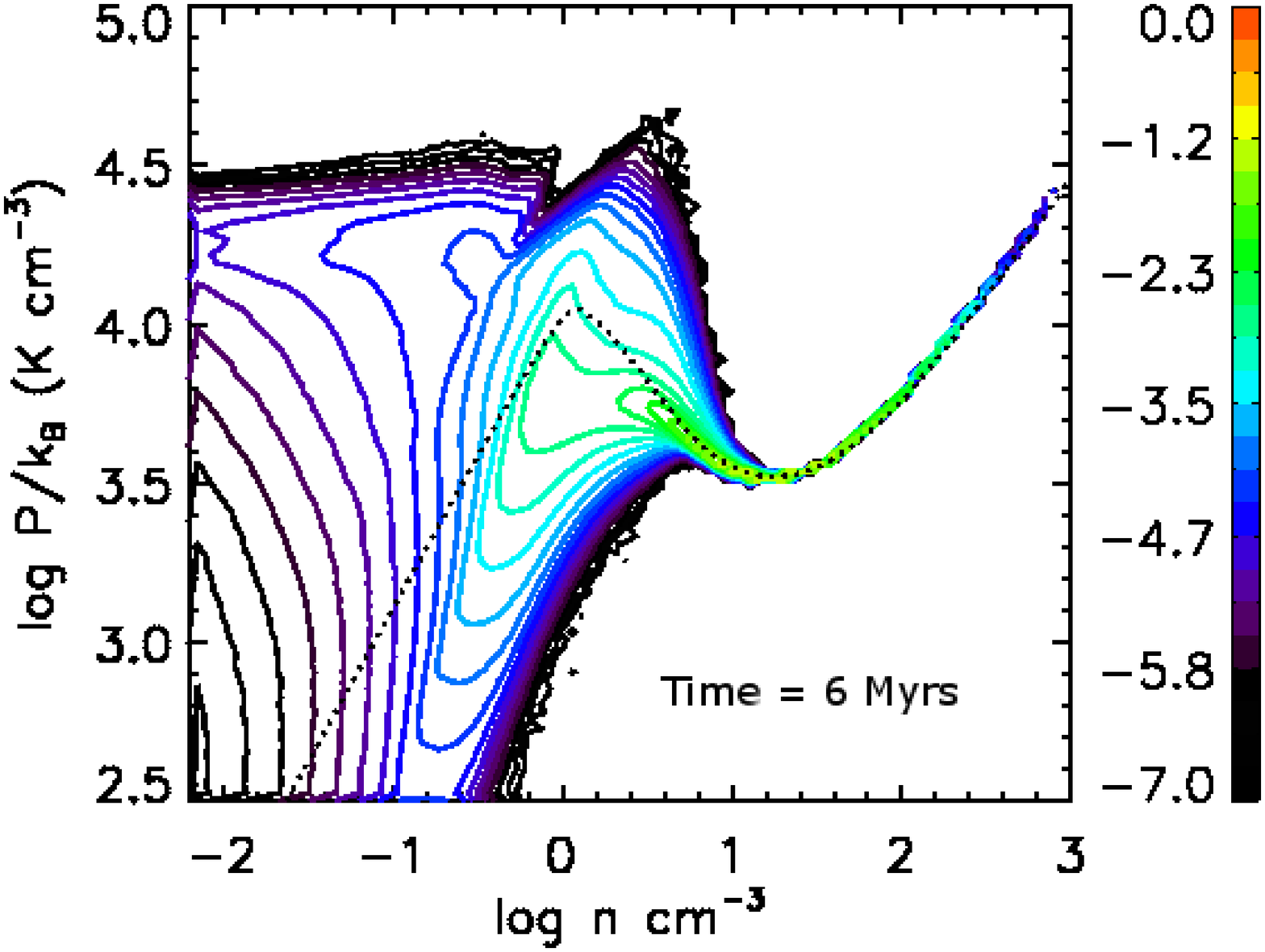}
    \includegraphics[width=0.33\linewidth]{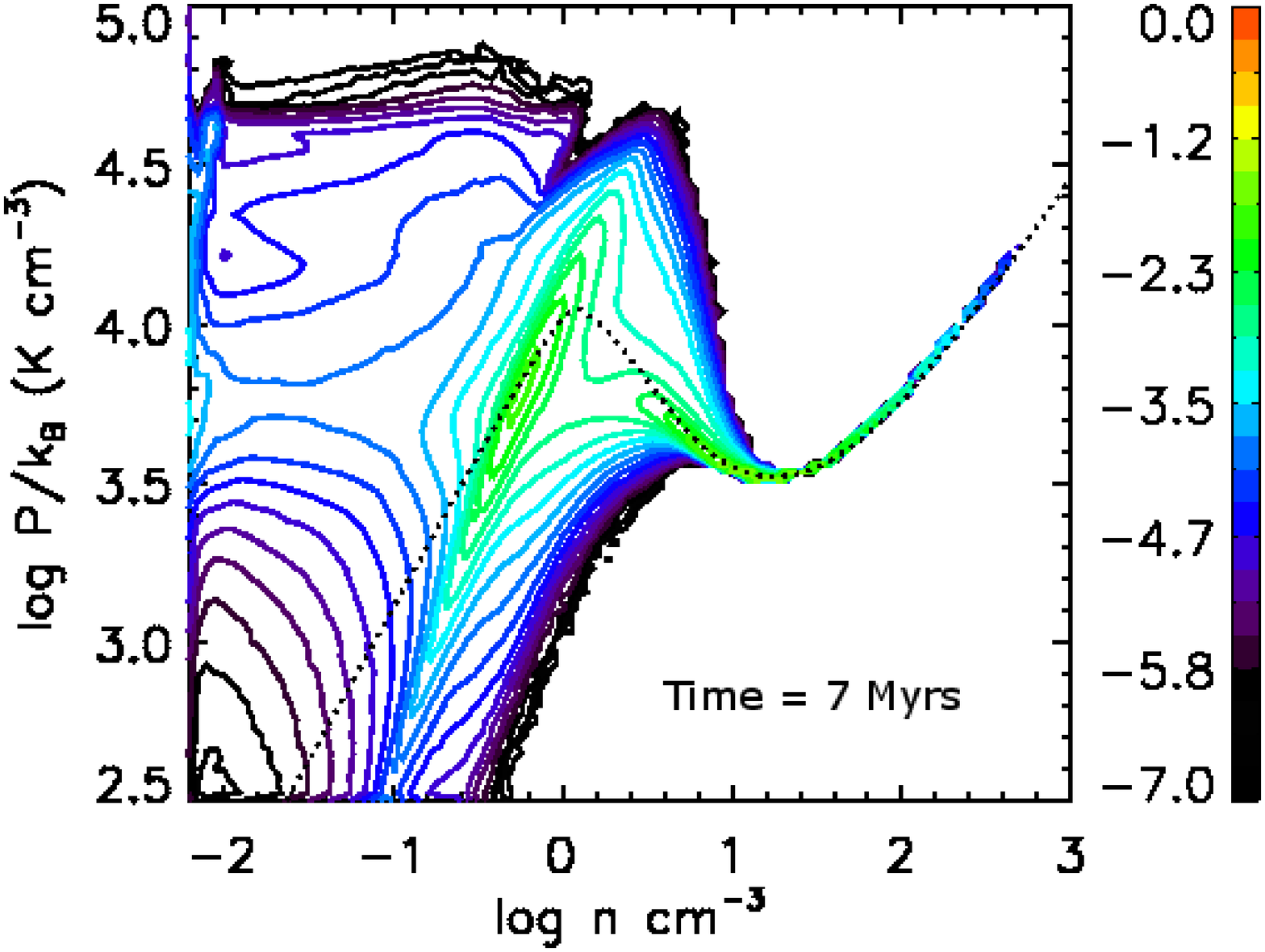}
    \caption{Two-dimensional mass histogram in pressure and density bins for runs \textit{hyd} (left; run \textit{hydg} is almost identical), 
\textit{mhd1r} (middle), and \textit{mhd1t} (right).
Top panels show an early phase (before the collision). Bottom panels show a late phase of the collision at the end of the simulation. 
The color  bars show the mass fraction in each bin and the dotted line indicates the cooling-heating equilibrium curve 
from the cooling function implemented in the code.}
     \label{TI}
\end{figure*}


\section{Introduction}

Supershells are large-scale shocks with radii larger than a hundred parsecs \citep{Heiles_1979,Tenorio-Tagle_1988}, 
commonly formed around clusters of young stars due to their very violent 
radiative and mechanical feedback \citep{McCray_Snow_1979,Bruhweiler_1980,McClure_2002}.
This powerful injection of momentum on large scales makes supershells an important driver of interstellar turbulence
\citep{Elmegreen_Scalo_2004,deAvillez_2007,Joung_2009} and a possible trigger for the formation of molecular clouds 
\citep{Ehlerova_1997,Ehlerova_2002,Dawson_11_1,Dawson_2013_rev}.

Supershell collision sites are also common, and they have often been associated with an enhanced molecular gas fraction 
compared to isolated shells \citep{Yamaguchi_2001,Preibisch_2002,Dawson_2013_rev, Fujii_2014,Dawson_15}.
This evokes the "colliding flows" scenario for molecular cloud formation
\citep{Hartmann_2001,Audit_2005,Heitsch_2006,VS_2007,Hennebelle_2008} according to which, 
when two supersonic flows collide, even small perturbations can trigger fluid instabilities that generate dense and turbulent gas.

A number of interesting questions arise in this situation, such as  how much dense gas forms from 
the fragmentation of an isolated supershell and how much is formed (or destroyed) by a supershell collision? 
In general, what is the signature of the supershell passage or of a collision in the kinematics and the structure of the surrounding interstellar medium (ISM)?
Answering these questions can significantly improve our understanding of an important driver of turbulence in galaxies,
and these questions are the main subject of this work.  Our particular focus is the role of the galactic magnetic field.

The idea that the condensed gas in large-scale shocks fragments
and causes the formation of molecular clouds is supported by observations of excess dense gas 
or young stars around superbubbles with respect to the field \citep{Walborn_1992,Oey_1998,Oey_2005,Dawson_11_1,Dawson_11_2, Dawson_13,Ehlerova_2016}.

From a theoretical point of view, it has been argued that large, dense gaseous fragments can form around superbubbles 
owing to a gravitational instability of the shell \citep{McCray_Kafatos_1987,Ehlerova_2002,Elmegreen_2002,Dale_2009,Wunsch_2010}.  
However, although the fragments on the shell should eventually become gravitationally unstable, 
dynamical instabilities are able to cause fragmentation earlier.  
\citet{Vishniac_83} performed a linear analysis of the dynamical stability of decelerating spherical shocks. 
His analysis included the effects of self-gravity, but the possible effect of a magnetic
field was only discussed qualitatively. 
The main conclusion there was that, for a pressure-confined shell, the gravitational instability of a given scale only becomes comparable to its dynamical instability if the original medium was already gravitationally unstable on that scale,  
and that an ordered magnetic field can stabilize the shell by making it thicker.

\citet{MacLow_1988} and \citet{MacLow_1989} presented some of the first numerical studies of superbubble dynamics, focusing mainly 
on their confinement in the galactic disk and their Rayleigh-Taylor stability (for a detailed description of this instability see following section). In their simulations the structure of the shocks 
was significantly affected by radiative cooling and by vertical acceleration out of the stratified galactic disk.
\citet{Tomisaka_1990} examined the case of a superbubble expanding in a magnetized environment numerically, and found that the Galactic magnetic
field was enough to prevent superbubbles below a certain luminosity from breaking out of the Galactic disk vertically. In a follow-up paper, \citet{Tomisaka_1992}
found that a superbubble would occupy up to 50\% less volume in a magnetized medium with respect to an unmagnetized 
environment.

At the same time, \citet{Ferriere_1991} studied the evolution of a superbubble in a uniform magnetic field by evolving the inner and outer radii
and showed that the shell tended to be thicker where the magnetic flux was larger.
This indicates that the magnetic field can indeed alter the stability properties of the shell through
affecting its thickness, but additional effects remain to be investigated.
More recently, \citet{Stil_2009} studied the evolution of superbubbles in homogeneous and stratified magnetized media.
As in previous studies, these authors found a significant elongation of the bubbles and a deceleration of the shells with increasing magnetic field strength.

So far, however, hardly any of the studies dealing with superbubble dynamics have included both the turbulent motions of the galactic gas and the turbulent component of the magnetic field.
Part of this work investigates the role of the mean and turbulent component of the galactic magnetic field
on the expansion and structure of the shocks.  The rest of this work deals with supershell collisions.

The nonlinear dynamics of the collision between two superbubbles was investigated by \citet{Ntormousi_2011}.
These authors found that the shells became dynamically unstable early in their evolution 
and that the collision created elongated clumps with high internal velocities due to large-scale shear. 
Their simulations were carried out in two dimensions, which allowed for very high resolution, but at the same
time limited the models from including magnetic fields or full turbulent dynamics.

Three-dimensional hydrodynamical simulations of superbubbles 
evolving and colliding in the warm diffuse phase of the ISM were recently presented in \cite{Dawson_15} 
as part of a comparison with observations of a molecular cloud being compressed by two supershells.
The two simulations used in that comparison are also part of this work.
One of the conclusions in that work was that the shock collision in the simulations did not
produce an enhancement in dense gas able to explain the observed molecular gas.
This left room for a scenario in which the observed cloud is a result of a pre-existing overdensity between the two shells,
but the parameter space for the shell collision was far from adequately explored.

In this paper we study the expansion and fragmentation of two supershells until the point where they collide with each other.
This situation is essentially the integral scale of a turbulent cascade; therefore understanding the dynamics under realistic ISM conditions can
significantly improve our insight into the driving of turbulence.  
We look for characteristic signatures of the shock passage and collisions in the dense phase of the surrounding ISM. 
The novelty with respect to previous studies is that we include the effects of the galactic magnetic field. We show that, for the density regimes we consider,
self-gravity plays a minor role in the fragmentation process,
but the presence of a magnetic field completely alters the behavior of the models.

In Section \ref{insta} we briefly review the various fluid instabilities
that appear in the simulations.  In Section \ref{numerics} we give an overview of the numerical methods we employed and in 
Section \ref{evolution} we show the evolution of the shells in different models.
Section \ref{sec:filaments} describes the properties of the dense fragments formed on the shock surfaces.
Section \ref{discussion} summarizes our results and Section \ref{conclusions} concludes the paper.
 

\begin{figure}[h]
    \includegraphics[width=\linewidth]{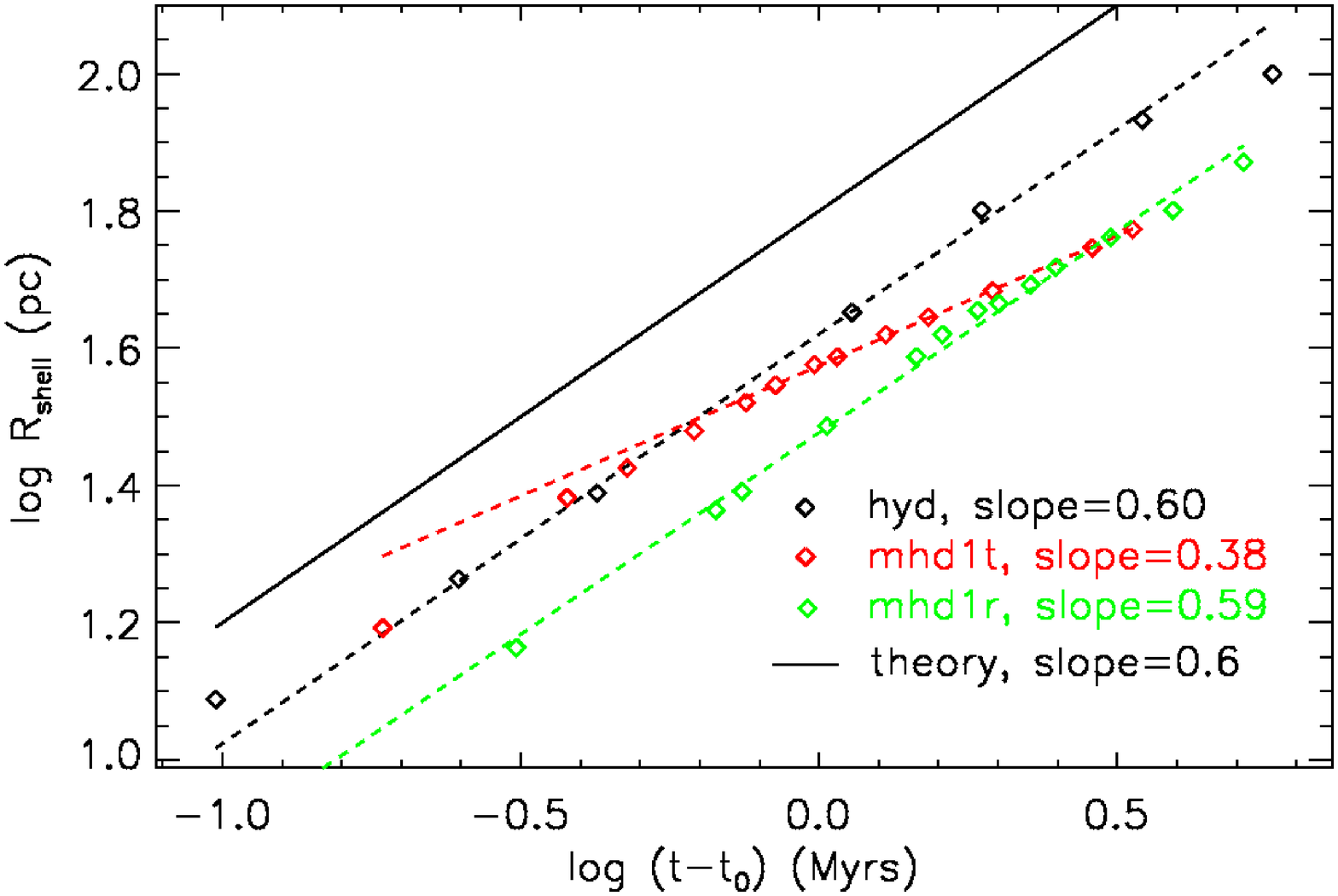}
    \caption{Logarithm of the average shell radius vs. logarithm of time for the different models.
Time is plotted relative to the first available snapshot for each run at about 0.2~Myrs.
The dashed lines are linear fits with the slope mentioned on the bottom right of the plot. 
The analytical radius-time relation, $R(t)\propto t^{3/5}$, almost coincides with the black dashed line;
the offset with respect to the dashed lines was introduced for illustration purposes.}
     \label{shells_radius_time}
\end{figure}
\begin{figure}[h]
    \includegraphics[width=\linewidth]{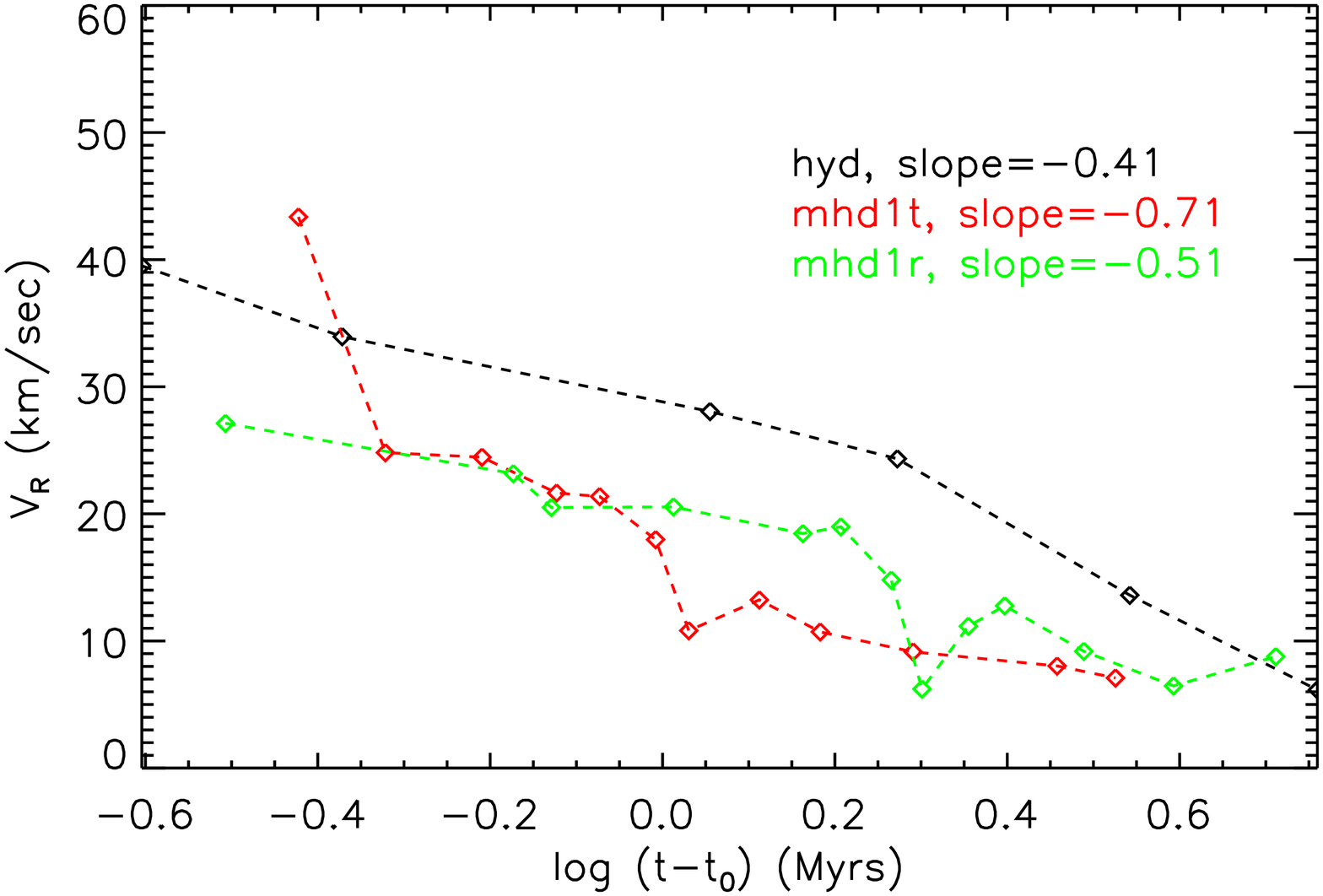}
    \caption{Average shell radial velocity (calculated as the relative change in radius) vs. the logarithm of time for the three models \textit{hyd}, \textit{mhd1r,} and \textit{mhd1t}.
Time is plotted relative to the first available snapshot for each run at about 0.2~Myrs.
The slopes of the linear fits are mentioned on the top right of the plot.}
     \label{shells_velocity_time}
\end{figure}

\section{Outline of relevant fluid instabilities}
\label{insta}

We consider a spherical shock that is subject to cooling, so it becomes very thin compared
to its radius, and is steadily decelerating radially. This initially ordered flow is subject to
a number of fluid instabilities, which can eventually lead to turbulence 
as they evolve into the nonlinear regime.

\subsection{The Vishniac instability}

In the previous section we mentioned the instability of a decelerating, thin, spherical 
shock, presented by \citet{Vishniac_83}. This instability is triggered by the different
natures of pressure confining the shell.  At the interior there is thermal pressure, which is 
isotropic, while the exterior is bounded by ram pressure, which is always directed normal to the shock surface.
It has been shown that for an adiabatic exponent larger than unity, a small ripple on the shock surface tends to grow under the influence 
of these two pressures, transferring momentum from the radial to the tangential directions.  
This bending mode of the shock can lead to a significant deformation of the surface, or even a break up of the shell as it becomes nonlinear.

In the linear regime the growth rate is proportional to the square root of the perturbation wave number k,
\begin{equation}
  \sigma_V = \left(P_i k^2 \alpha \slash \Sigma\right)^{1/4}
,\end{equation}
where $P_i$ is the thermal pressure driving the shock expansion, $\alpha$ is the amplitude of the (negative) 
shock acceleration, and $\Sigma$ is the column density from the inner to the outer shock surface.

\subsection{The Rayleigh-Taylor instability}

A flow instability that has been observed in many astrophysical environments
is the Rayleigh-Taylor (RT) instability of two superimposed fluids.  If the gravitational (or another) acceleration
is directed opposite to the density gradient of the flow, then the flow is unstable and parcels of dense fluid move into the 
diffuse fluid, eventually saturating into a mixing layer.  

The growth rate of the RT instability is 
\begin{equation}
   \sigma_{RT} = \sqrt{A g k}
 \label{gr_rt}
,\end{equation}
where, if $\rho_1$ is the density of the heavier fluid and $\rho_2$ is the density of
the lighter fluid, $A = (\rho_1-\rho_2) \slash  (\rho_1+\rho_2)$, $g$ is the magnitude of the acceleration
 and $k$ is the spatial wave number of the mode.

A magnetic field perpendicular to the density gradient can stabilize modes smaller than 
a critical wavelength $\lambda_c$,
\begin{equation}
\lambda_c = \frac{B^2}{g(\rho_1-\rho_2)}
,\end{equation}
where B the magnitude of the magnetic field.
A description of the instability and a derivation of these growth rates can be found in \citet{chandra}.

\subsection{The thermal instability}

The thermal instability (TI) of interstellar gases was first described by \citet{Field_1965} and is responsible for creating the cold and warm phases of
the atomic ISM. Put in a simplified way, this instability arises when a perturbation (condensation, change in temperature) 
in a fluid moves against the gradient of the cooling function $\Lambda$ with respect to the corresponding variable (density, temperature).  
Then the perturbed fluid parcel undergoes runaway 
cooling or heating, until reaching a stable point of the cooling function. 
In the context of a thin decelerating shock, or converging flows, the TI can be a consequence of 
local condensations or changes in pressure.

Under ISM conditions, the TI grows in an isobaric mode, which means that the smallest
allowed perturbations grow first.  If $\rho$ is the fluid density, condensations smaller than the Field length
\begin{equation}
 \lambda_F = \left(\frac{\kappa T}{\rho^2 \Lambda}\right)^{1/2} 
,\end{equation}
which is controlled by the coefficient of thermal conductivity, $\kappa$, of the fluid, cannot grow \citep{Field_1965,KI_2004}. 

\subsection{The Kelvin-Helmholtz instability}

The instability of
two superimposed fluids of opposite horizontal velocities is another instability that is not directly associated with a decelerating shock, but can 
result from perturbations in this environment.  A perturbation in the vertical velocity then tends to grow, forming characteristic eddy-like features along the shear layer and eventually
leading to a mixing layer in the nonlinear regime.

The Kelvin-Helmholtz (KH) instability, similar to the RT instability, in the linear regime 
grows fastest at the smallest scales, with a growth rate  
\begin{equation}
  \sigma_{KH} = k\Delta U
,\end{equation} 
where $\Delta U$ is the difference in horizontal velocity of the two fluids.
A surface tension between the two fluids or equivalently, a magnetic field perpendicular to the velocity gradient  
stabilizes modes for which the velocity difference is smaller than the Alfv\'{e}n speed \citep{chandra}.


\section{Numerical code, initial and boundary conditions}
\label{numerics}

\subsection{Code and additional model implementations}

We model two colliding superbubbles in three dimensions 
using the magnetohydrodynamical (MHD) code RAMSES \citep{Teyssier02,Fromang_2006},  which is
suitably adapted to account for typical ISM heating and cooling and to simulate the feedback from young stellar populations.  

The superbubbles are created by thermal and kinetic feedback from OB associations, 
which is approximated in the code by equally distributing the total 
thermal energy and mass output from 30 stars among a group of cells inside a spherical region of 5 pc radius.  
The feedback masses and energies have been calculated and provided to us by 
\cite{Voss09} as the average mass and energy output from a typical galactic OB association, 
including stellar winds and supernovae.  
Both the wind module and the cooling and heating function are the same as in \citet{Ntormousi_2011} and \citet{Fierlinger12}.

Radiative effects are not simulated in our models, although in the initial stages of stellar
evolution they can be dynamically very important \citep{Geen_15,Ngoumou_2015}.  
One-dimensional test models with radiation, however, showed that the shock front is always ahead of the radiation front  for the
timescales we are studying.

We need to model the cooling and heating processes in the ISM to simulate the formation of a two-phase medium.
Heating comes from photoionization of the gas and from the photoelectric effect on dust grains.  
Cooling is due to atomic lines, i.e., mainly carbon and oxygen.
\citet{Wolfire95} provide a very detailed description of all these processes. 
We used the rates from that work here in a tabulated form as a function of density and temperature for a gas of solar metallicity.
\subsection{Initial conditions}

To study the effects of gravity and magnetic fields on the structure and evolution of the supershells 
we test three types of models: pure hydrodynamics, hydrodynamics with self-gravity, and 
non-self-gravitating MHD.  
The different models are summarized in Table \ref{tab_models} of the next section.
The power of the feedback (30 OB stars) and the size of the box (200~pc) do not vary between the different models.

Previous calculations \citep{Ntormousi_2011}, although two-dimensional, showed that modeling the 
supershell expansion in a turbulent environment produces more realistic dense structure around the bubbles, 
so we evolve all the models in a turbulent diffuse ISM.  

The turbulence is created following \citet{MacLow_1999}:
We first introduce random phases to four wave numbers ($k$=1-4) in Fourier space and then 
take an inverse Fourier transform to create the three components of the velocity.  
This velocity field is applied to a box of initially homogeneous density (and uniform magnetic field, for the 
magnetized models) and is integrated 
until the density-weighted power spectra of the turbulence reach Kolmogorov behavior.  

The turbulence driving process is exactly the same for the magnetized runs except that we start 
with a magnetic field along one direction that is either parallel or perpendicular to the collision axis. 
Because of the conservation of magnetic flux, even when the turbulence
has become almost completely isotropic, there is still a mean magnetic field component along that direction.   
This is consistent with a representation of the Galactic mean magnetic field.  
The turbulence fluctuates with a sonic rms Mach number around a mean number density of 1 cm$^{-3}$.  
The mean temperature is set to the equilibrium of the cooling curve for this density, which is 8000 K. 

We are simulating a cube of a 200 pc$^3$ physical size at a uniform resolution of 512$^3$ grid points.  
In the run including self-gravity we introduced one AMR level to resolve the Jeans length of the densest gas.
The feedback areas are placed on either side of the box along the x-axis in the hydro runs and along the z-axis in the MHD runs.   
The boundary conditions are open (zero gradients).

\section{Model evolution}
\label{evolution}

Table \ref{tab_models} summarizes the different models included in this work.
We first present the hydrodynamical models and then discuss the effects of adding a magnetic field.  

\begin{table}
\caption{Summary of simulation parameters. 
The magnetic field strength and orientation in the second and third columns refer to the initial configuration of the mean field
with respect to the collision axis of the bubbles.}
  \begin{tabular*}{\linewidth}{@{\extracolsep{\fill}}llll}
    \hline
    \textbf{Name}  &  \textbf{B$_{in}$}  & \textbf{B$_{in}$ (orientation)} & \textbf{Self-gravity}  \\ 
    \hline
    \hline
     hyd     & 0          & -   &  no   \\
     hydg   & 0          & -   &  yes  \\
     mhd1r  & 5~$\mu$G   & parallel     &  no  \\ 
     mhd1t  & 5~$\mu$G   & tangential     &  no  \\ 
     single1l & 5~$\mu$G   &  -   &  no  \\
     single10l & 1.5~$\mu$G   &  -   &  no \\
    \hline
  \end{tabular*}
\label{tab_models}
\end{table}
%
\begin{figure*}[th!]
   \includegraphics[width=0.32\linewidth]{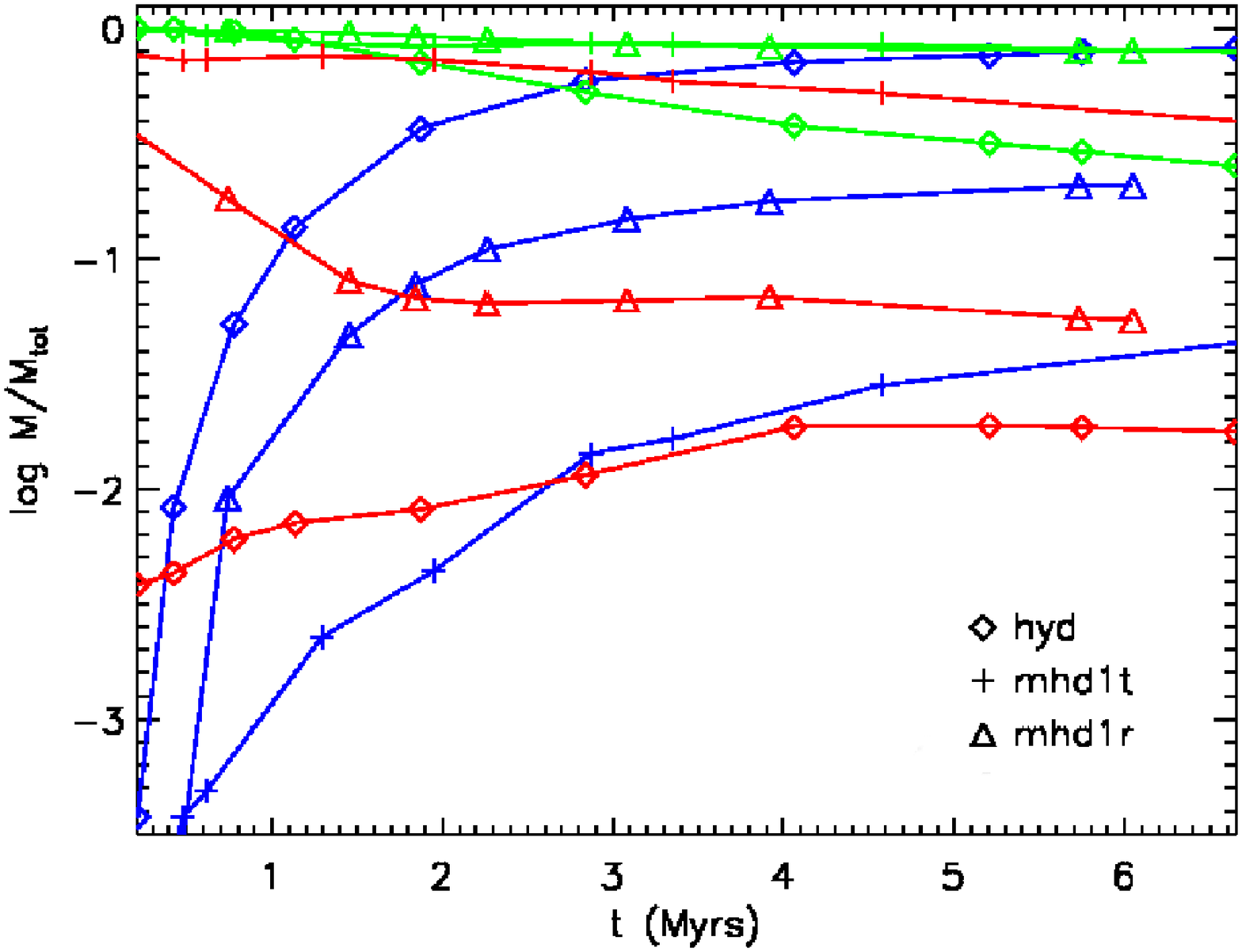}
   \includegraphics[width=0.32\linewidth]{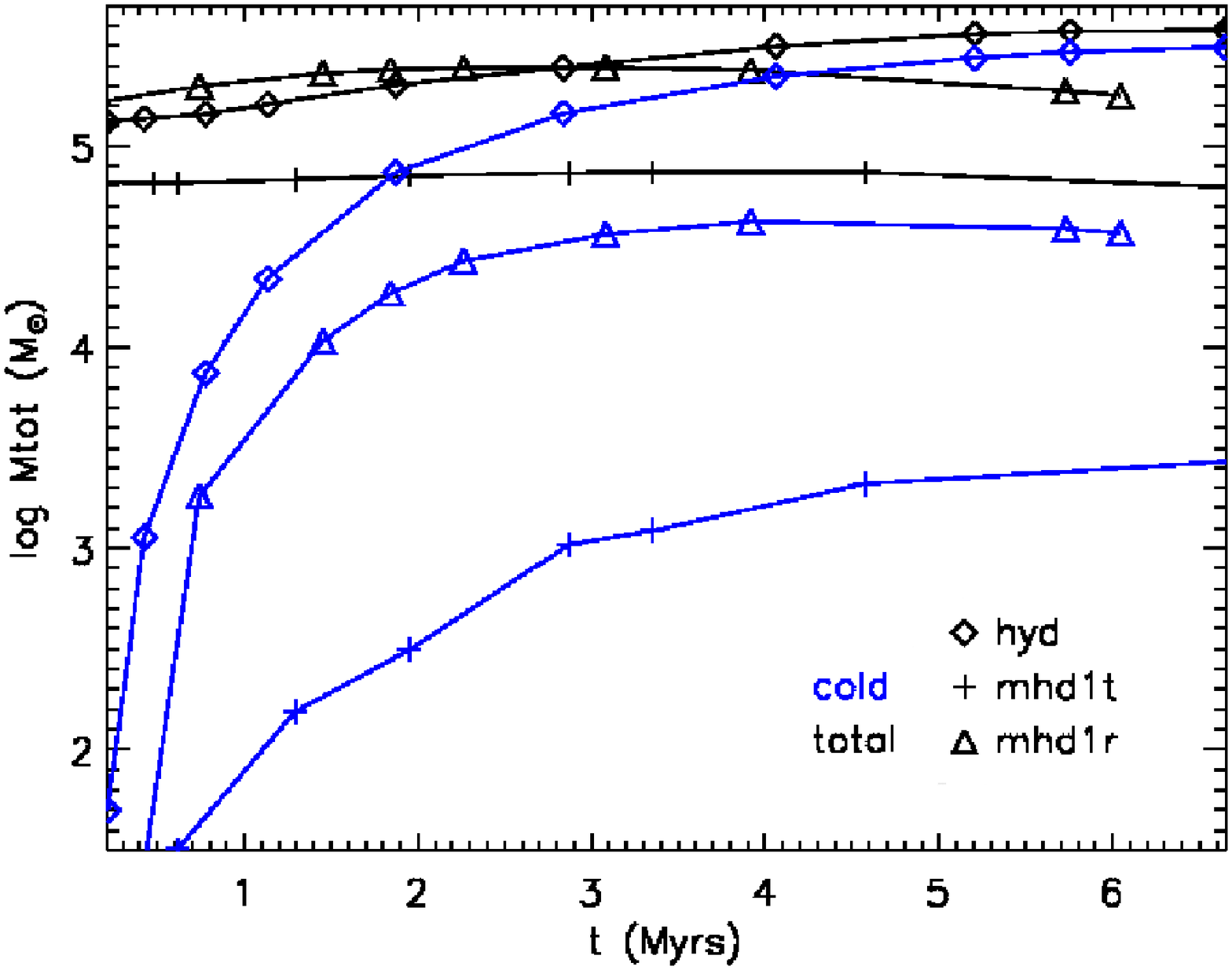}
   \includegraphics[width=0.32\linewidth]{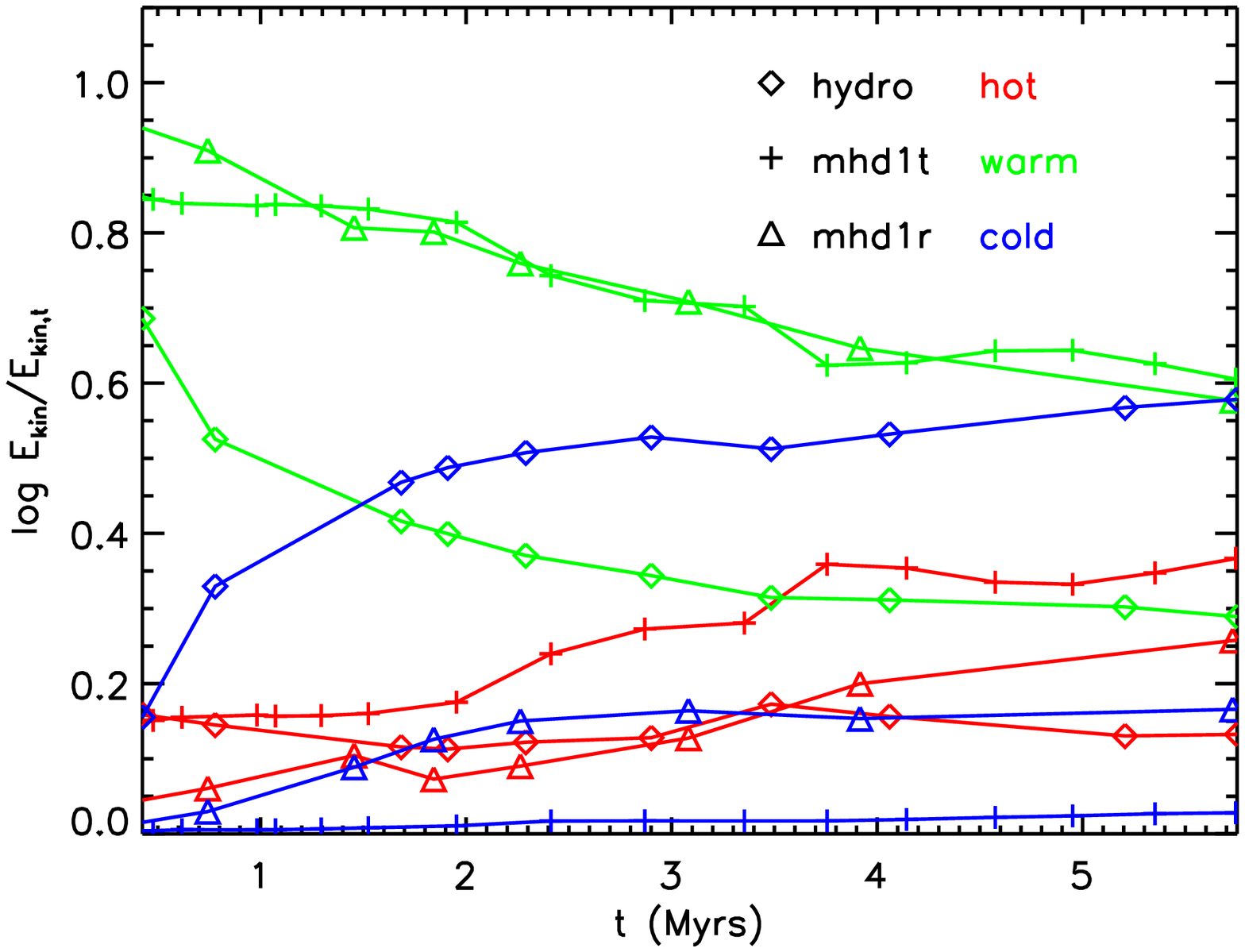}
   \caption{Logarithm of gas mass fractions (left), absolute gas masses (middle), and kinetic energy fractions (right) per gas phase as a function of time. 
 The diamond shapes correspond to run \textit{hyd}, the crosses correspond to run \textit{mhdr,} and the triangles to run \textit{mhdt}.  Blue indicates temperatures below 100~K, green temperatures $100<T\le 10000$~K, and red temperatures above 10000~K. The black curves in the middle panel show the total mass in the computational box.}
   \label{gas_phases}   
\end{figure*}

\begin{figure}[h!]
   \includegraphics[width=\linewidth]{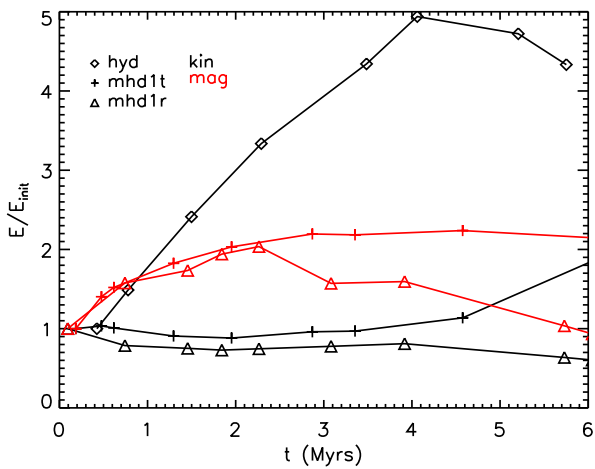}
   \caption{Total kinetic and magnetic energies in the box as a function of time, shown with respect to the initial values. 
The diamond shapes correspond to run \textit{hyd}, the
crosses correspond to run \textit{mhd1r,} and the triangles to run \textit{mhd1t}.   Black indicates the total kinetic energy and red the total magnetic energy.}
   \label{ener_time}
\end{figure}

\subsection{Evolution with hydrodynamics and self-gravity}

%

The two nonmagnetized models (\textit{hyd} and \textit{hydg} in Table \ref{tab_models}) were used in the comparison with an observed molecular cloud between two superbubbles in \citet{Dawson_15}
and are similar to the two-dimensional hydrodynamical models presented in \citet{Ntormousi_2011}.
Both models simulate the bubble expansion and collision in a turbulent medium, but \textit{hydg} includes self-gravity.

Given the mean density of the environment (1 cm$^{-3}$) and within the timescales simulated (on average a total of 6.5~Myrs per model), we
do not expect gravity to be dynamically important for the supershell evolution.  
However, we included gravity in model \textit{hydg} because it could make a difference once dense structures formed on the supershell surfaces. 
The free-fall time ($t_{ff}=\left(\frac{3\pi}{32G\rho_0}\right)^{1/2}$) 
of gas with densities $n_H\simeq100$ cm$^{-3}$
is about 5 Myrs and drops to about 1 Myr for structures with $n_H\simeq1000$ cm$^{-3}$.
As we show in this paper, clumps of such densities only form at later stages of the hydrodynamical simulations.

Indeed, for most of the duration of run \textit{hydg} the results are identical to the case without self-gravity (run \textit{hyd}). In order to avoid repetition
 we only show plots for run \textit{hyd}.
Some deviations appear in the latest stages of the evolution and are presented in a following section, in which we discuss the properties of the dense gas. 

A three-dimensional view of the gas density in run \textit{hyd} is shown in the top panel of Fig. \ref{volume_rendering_all}, 
at 3.5 Myrs (left), and at 4.5 Myrs (right) into the evolution of the model, when the two bubbles have already started colliding.  
The color coding corresponds to the logarithm of the total gas density, with certain density ranges hidden to allow viewing the dense clumps more clearly.  
In particular, the visible gas is either very dilute wind gas in red, or gas above 50~cm$^{-3}$, which is our definition of dense gas.  
The arrows show the local magnetic field, normalized to the highest value in the box.
The shells are clearly fragmenting into clumps and, although some structure is present in the surrounding medium before the passage of the shocks 
due to the turbulence, the clumps located on the surfaces of the shocks are denser and larger.  
It is also worth mentioning that simulations of the same setup without the superbubbles produced almost no dense gas;
by dense gas, we mean gas with densities above 50 cm$^{-3}$, which would be classified as cold neutral medium (CNM) or a molecular cloud precursor.

In the top panel of Fig. \ref{shells_col_den_all} we show snapshots of the same simulation in column density.
It is clear that in this situation the column density map is a good approximation of the three-dimensional position and morphology of the shocks.
The signature of the collision is an increase in column density where the two shells meet.
Nonetheless, as pointed out in \citet{Dawson_15}, the increase is only roughly by a factor of two, a mere addition of the material swept up and condensed by the two shells.
This result remains unchanged when self-gravity is taken into account.

The shells appear to fragment early in the simulation, well before they collide.
As \citet{Vishniac_83} showed, a decelerating, spherical, thin shock is dynamically
unstable when the effective equation of state is steeper than isothermal.  
The surface of the shock is supported by thermal pressure on the inside and by ram pressure on the outside.
When there is a small perturbation on the shock surface, such as a ripple, 
the different natures of these pressures, one isotropic and one always normal to the surface,
cause a flow of material toward the tips.

Indeed, in our simulations the shocks are distorted by finger-like ripples over their entire surfaces
and at different scales.  However, these ripples do not grow indefinitely, and certainly not to the point of dynamically breaking up the shell.  
Consistent with what \citet{MacLow_Norman_1993} found in their simulations of a single mode of the Vishniac instability, the amplification of the perturbation eventually stops.  
In their interpretation, this happens when the internal shock motions become supersonic and are therefore damped very efficiently by cooling.

However, the Vishniac instability is not the only possible cause of fragmentation.
As we presented in Section \ref{insta}, a gas parcel can become thermally unstable when the heating and cooling processes 
in the fluid cause small condensations to cool and condense further.   
The signature of this instability is the creation of a multi-phase medium, in which the cold gas appears in the form of under-pressured clumps (i.e., \citealp{Walch_11}).
In our simulations, we expect the TI to be triggered inside the shocks by the Vishniac instability, as the gas condenses at the tips of the ripples. However, it can also act in the material between the shocks, as a result of the large-scale compression.

We can look for thermally unstable gas by estimating the mass fraction in different density and pressure bins. 
This is shown in Fig. \ref{TI}, in the form of two-dimensional histograms.
The top panel shows early stages of the simulations and
the bottom panel corresponds to the end of each simulation.
The black dotted curve shows the cooling-heating equilibrium curve we employed in the simulation.  

Initially, all the mass is placed around the equilibrium regime of the curve that corresponds to the warm neutral medium (WNM), with $n_H=1$ cm$^{-3}$, and $T=8000$~K.
However, as we see at the top left of Fig. \ref{TI}, after only 1.6~Myrs much of the gas has already moved from the WNM equilibrium into the thermally unstable regime and toward the CNM equilibrium  with $n_H\geq50$ cm$^{-3}$ and $T\leq100$~K.  
At the end of run \textit{hyd}, described by the bottom plot of the left panel, more material has moved to
the CNM equilibrium, marking the formation of a two-phase medium.  However, this gas is still under ram pressure from the shocks, so it 
moves toward even higher densities in search of pressure equilibrium. 
At all times, the hot cavity gas represents only a tiny fraction of the total mass with high pressure and low density.

Qualitatively speaking, we observe here the same overall behavior as in the two-dimensional models in \cite{Ntormousi_2011}.
Fluid instabilities, the most dynamical of which is the Vishniac instability, 
create rich structure within the shocks early in their evolution and TI causes them to fragment into small clumps.
In addition, the dense shock as a whole is decelerating, which
means that the acceleration is pointing from the denser to the more dilute gas.  This is a RT unstable configuration and it creates the characteristic finger-like structures of this instability
on the inner surface of the shock (not visible in the figures here).  These dense formations end up inside the hot wind and are quickly evaporated by the hot gas. 

\begin{figure*}[th!]
\centering
    \includegraphics[width=0.45\linewidth]{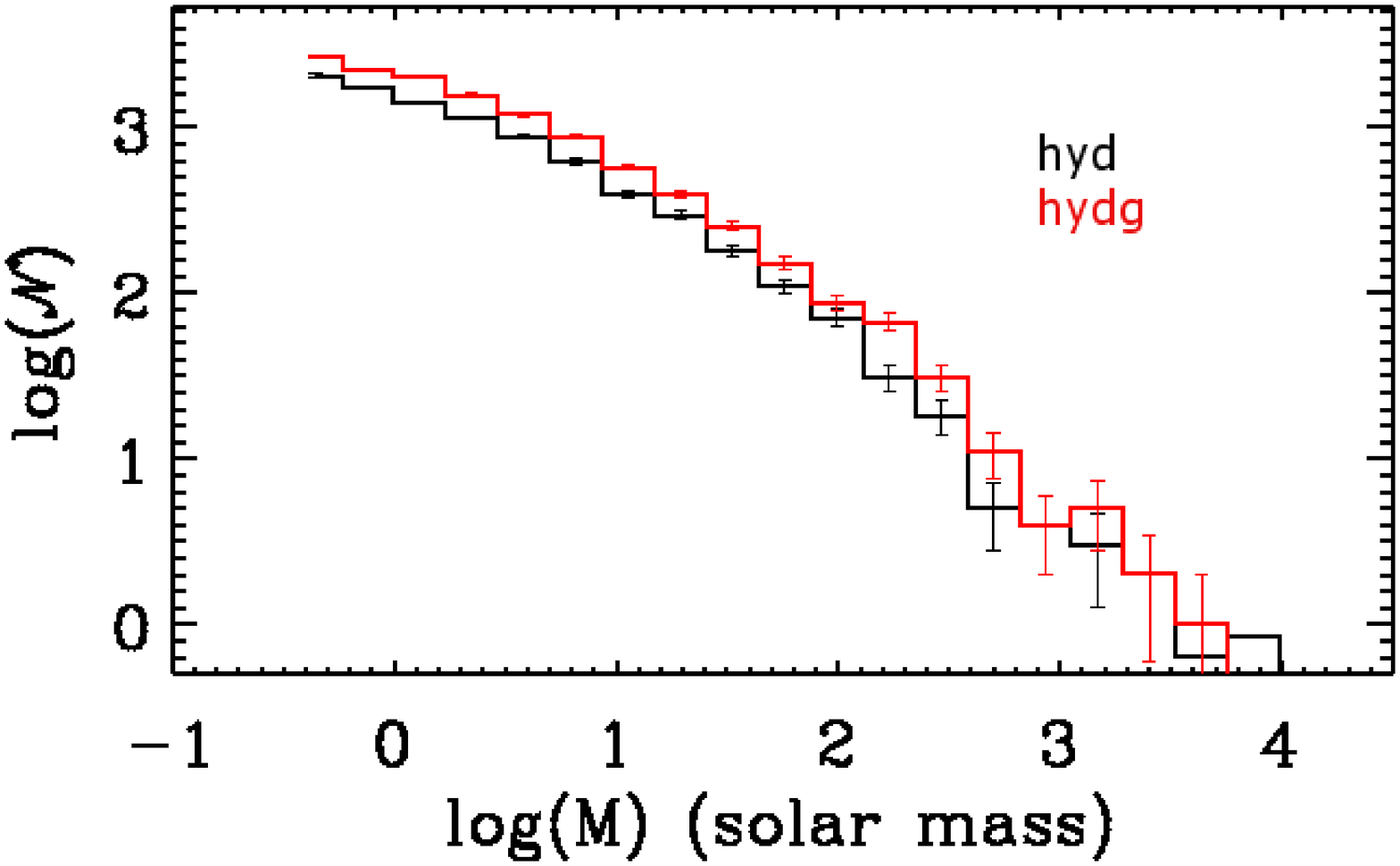}
    \includegraphics[width=0.45\linewidth]{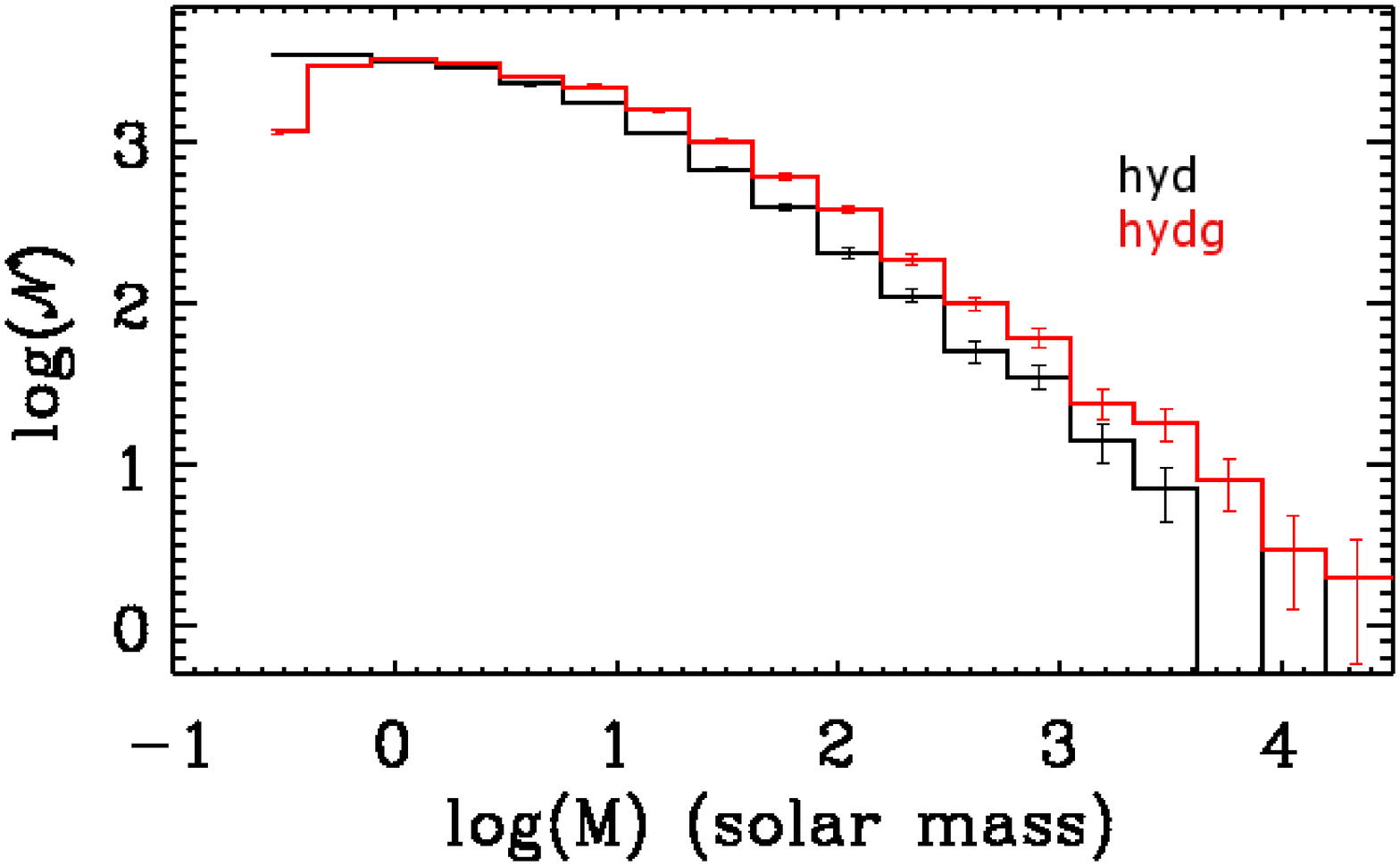}
    \includegraphics[width=0.45\linewidth]{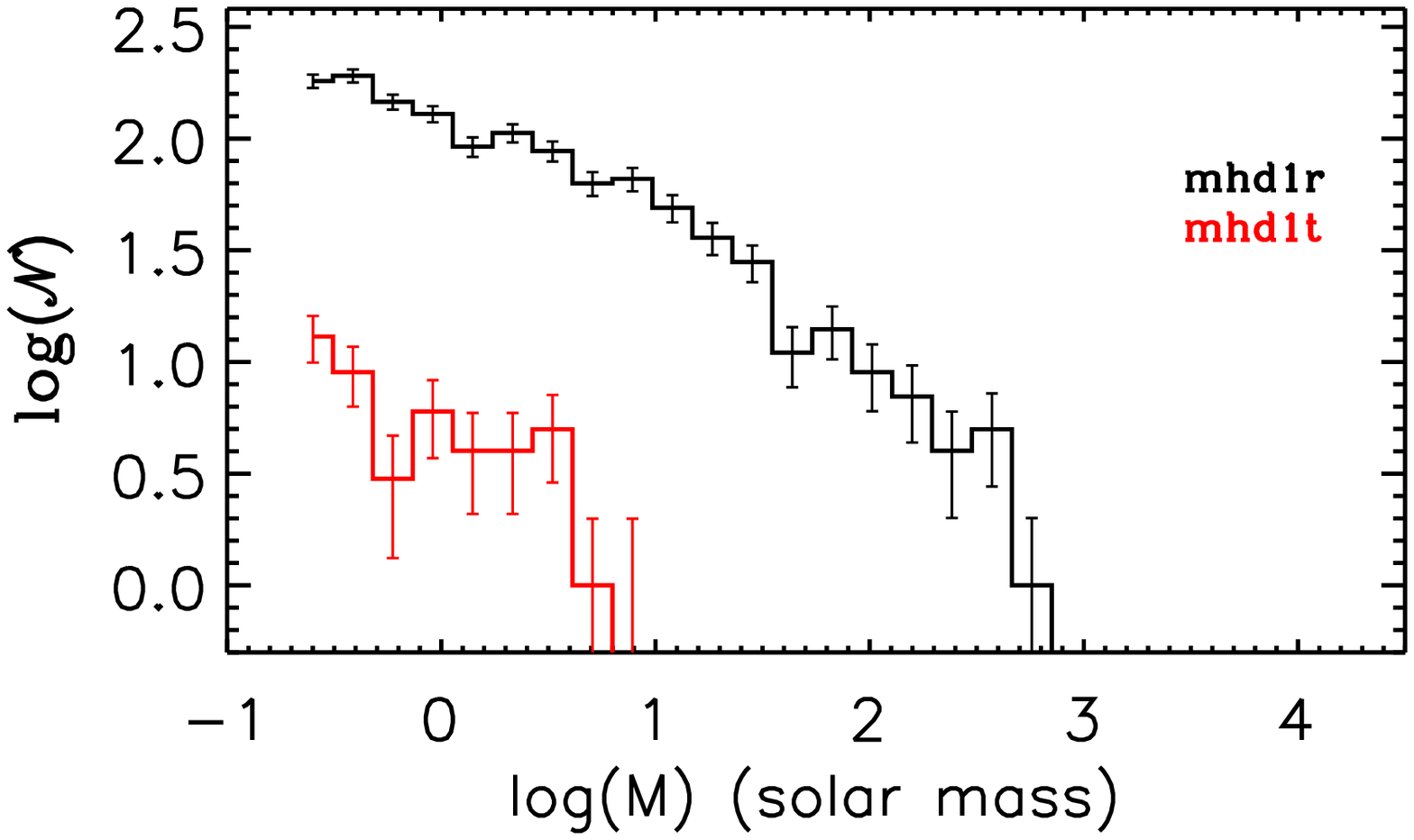}
    \includegraphics[width=0.45\linewidth]{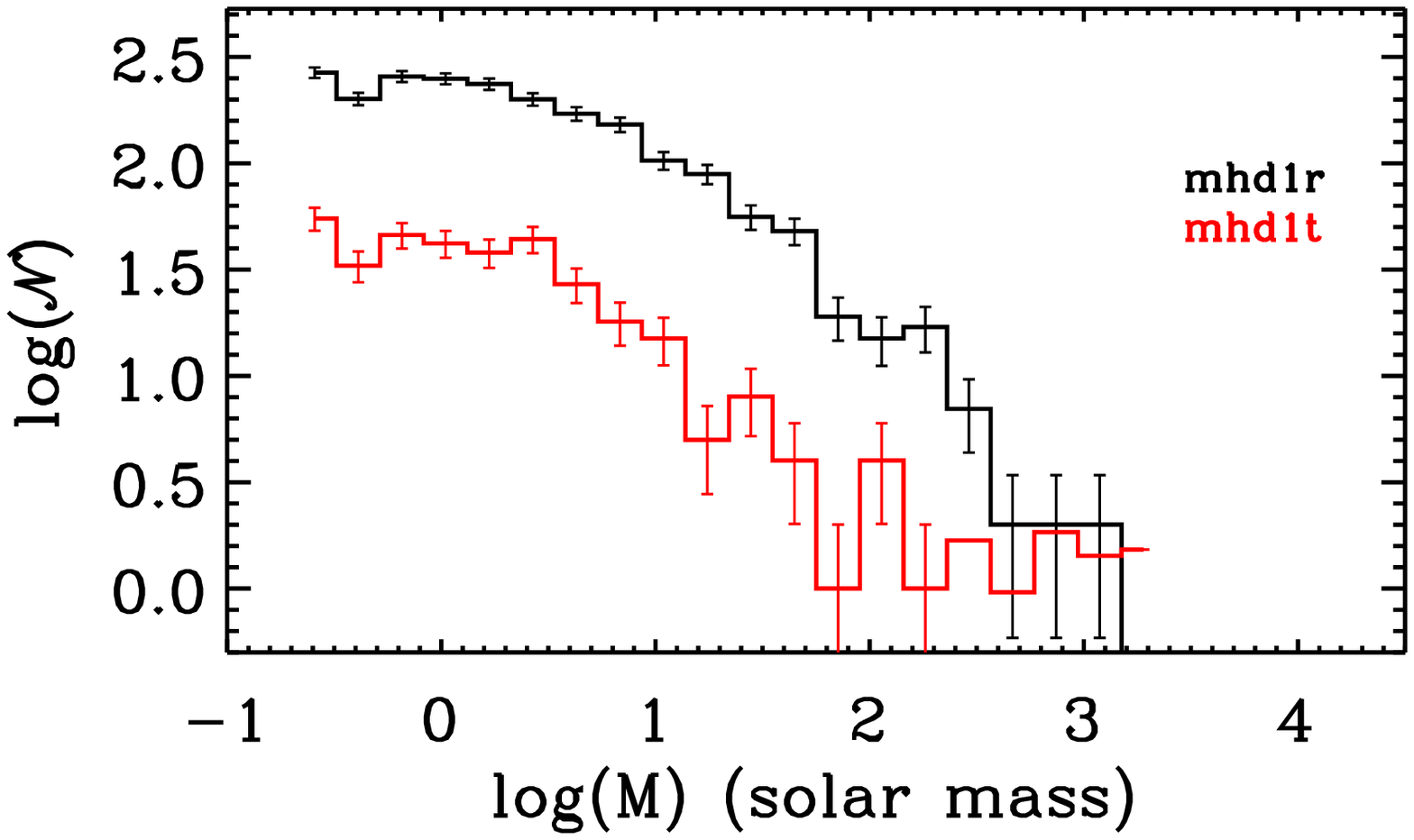}
    \caption{Distributions of the cold clump masses, defined as "friends-of-friends" groups
above a density threshold of 50 cm$^{-3}$, for different simulations. Top panel: \textit{hyd} (black histogram); \textit{hydg} (red histogram). 
Bottom panel: \textit{mhd1r} (black histogram); \textit{mhd1t} (red histogram).  Snapshots were taken at times t=2~Myrs (left) and t=6~Myrs (right).}
     \label{mass_histo}
\end{figure*}
\begin{figure*}[th!]
\centering
    \includegraphics[width=0.45\linewidth]{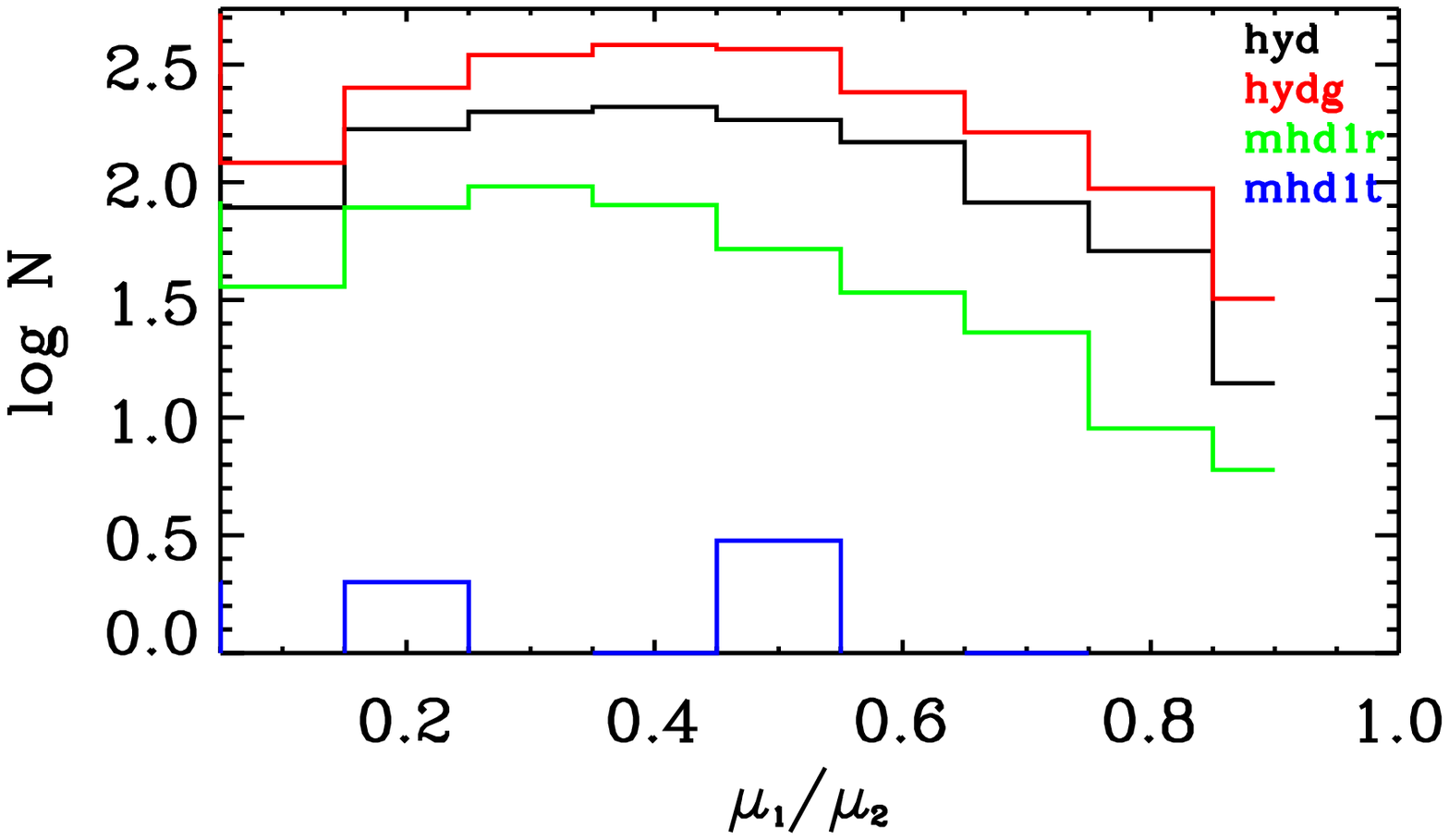}
    \includegraphics[width=0.45\linewidth]{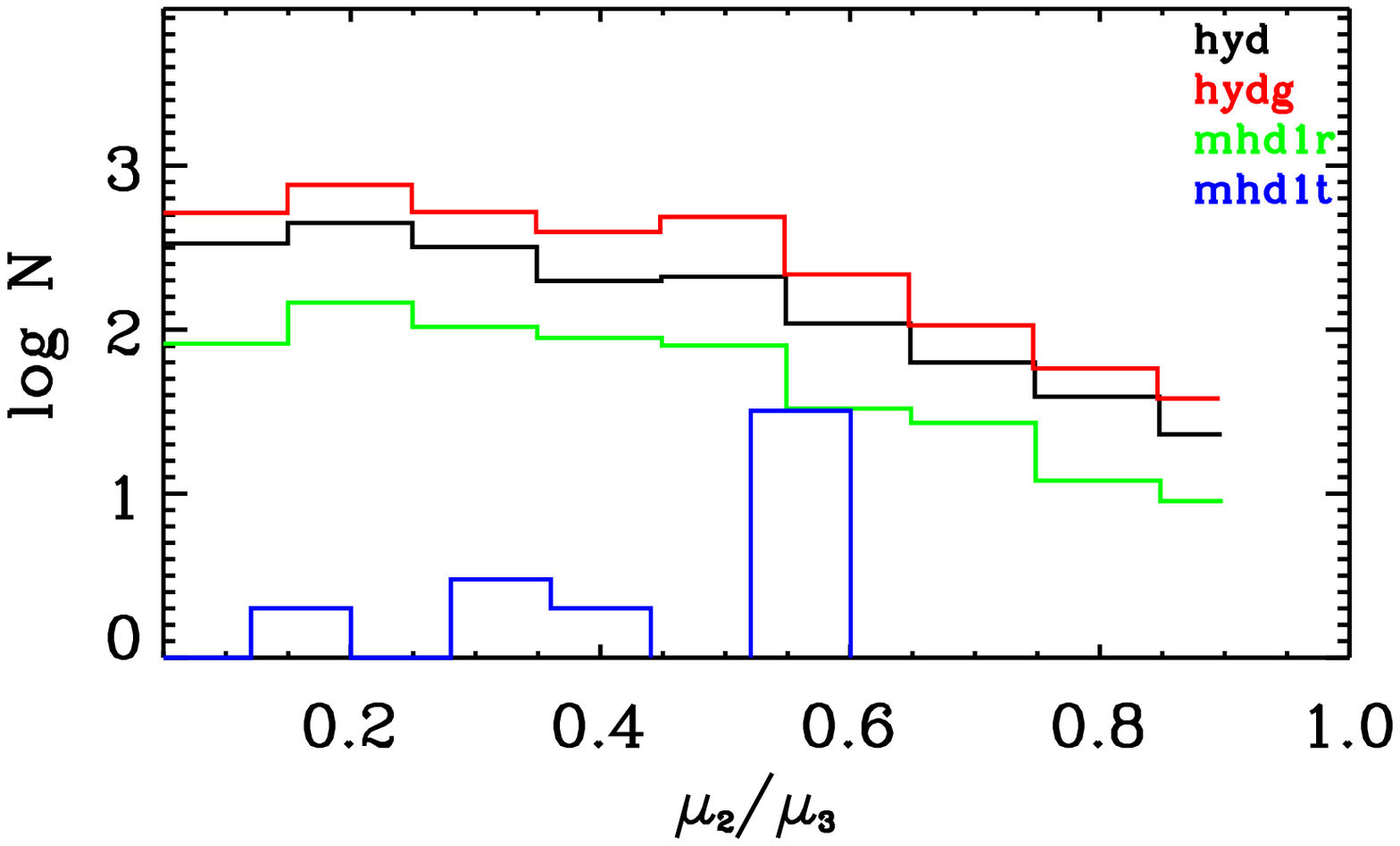}
    \includegraphics[width=0.45\linewidth]{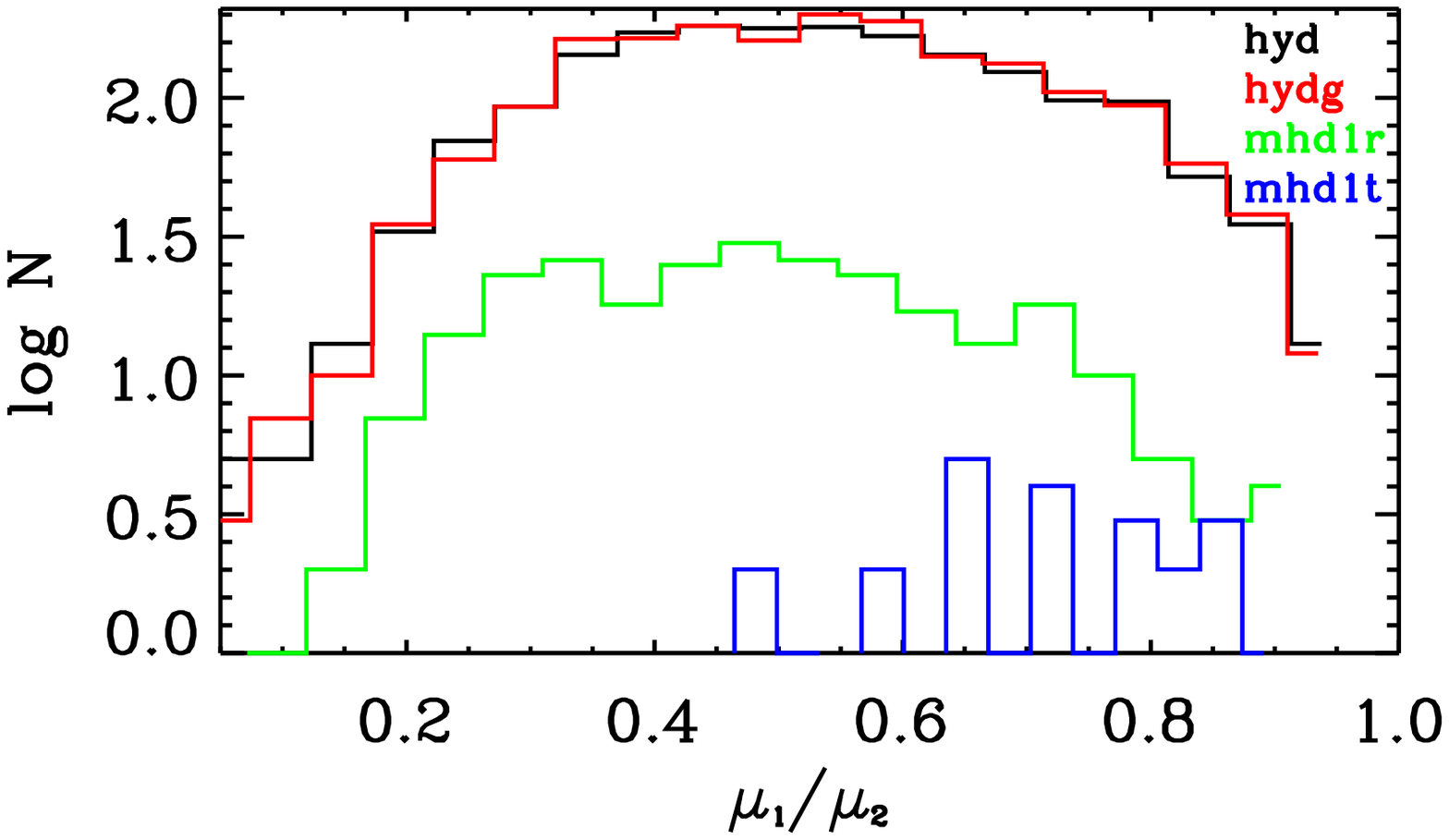}
    \includegraphics[width=0.45\linewidth]{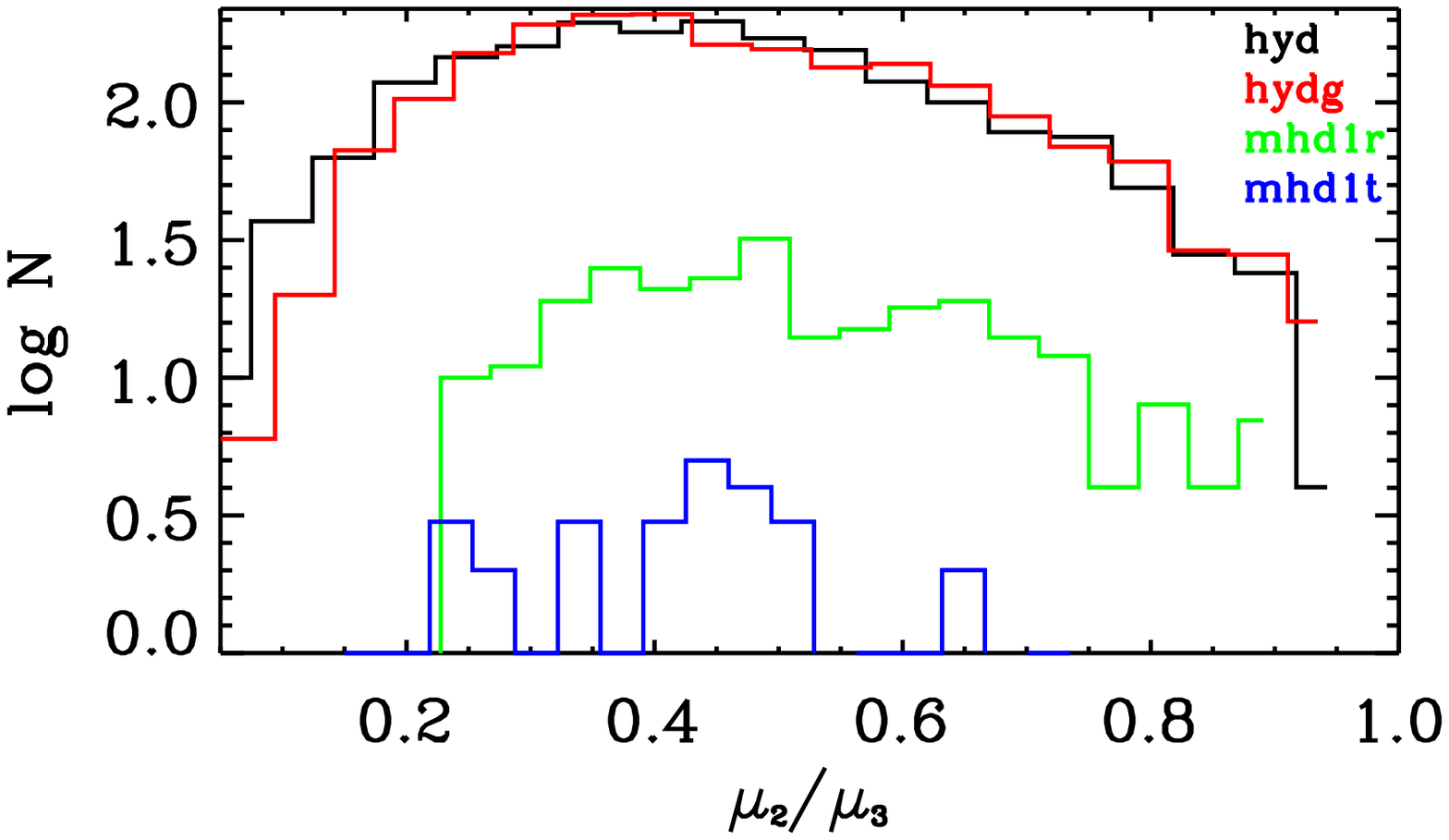}
    \caption{Clump axis ratio distribution in runs \textit{hyd}, \textit{hydg}, \textit{mhd1r,} and \textit{mhd1t} at about 2 Myrs (top), and at the end of each run (bottom). The end of runs \textit{hyd}, \textit{hydg,} and \textit{mhd1r} is at about 6 Myrs, and the end of run \textit{mhd1t} is at about 7~Myrs. 
The histograms on the left show the ratio of the shortest ($\mu_1$)
to the second longest ($\mu_2$) dimension of the structure, and those on the right show the ratio of the second longest 
to the longest ($\mu_3$) dimension of the structure. Clumps containing less than 10 cells are not shown.} 
     \label{axis_ratios}
\end{figure*}
%
\begin{figure*}[th!]
    \includegraphics[width=0.49\linewidth]{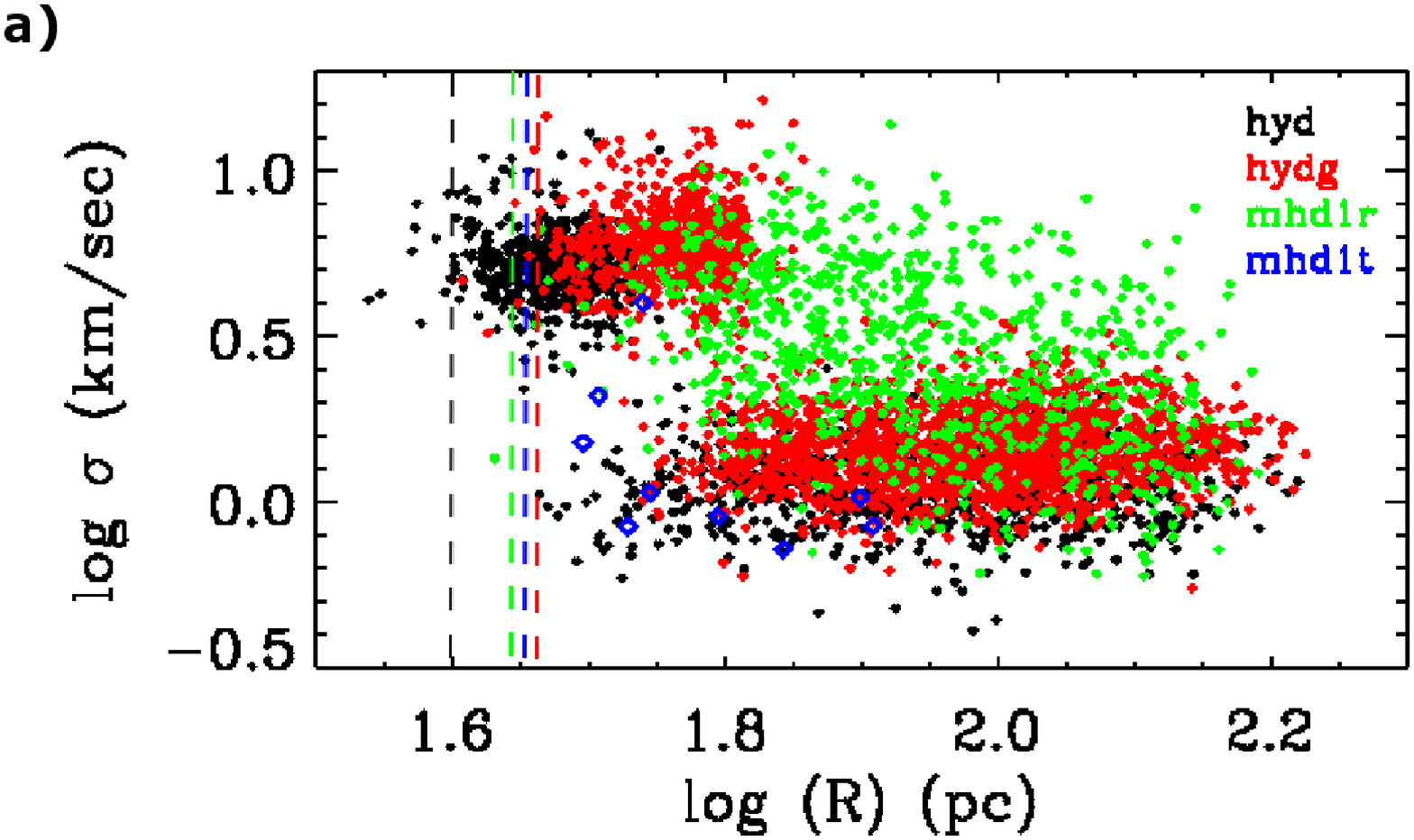}
    \includegraphics[width=0.49\linewidth]{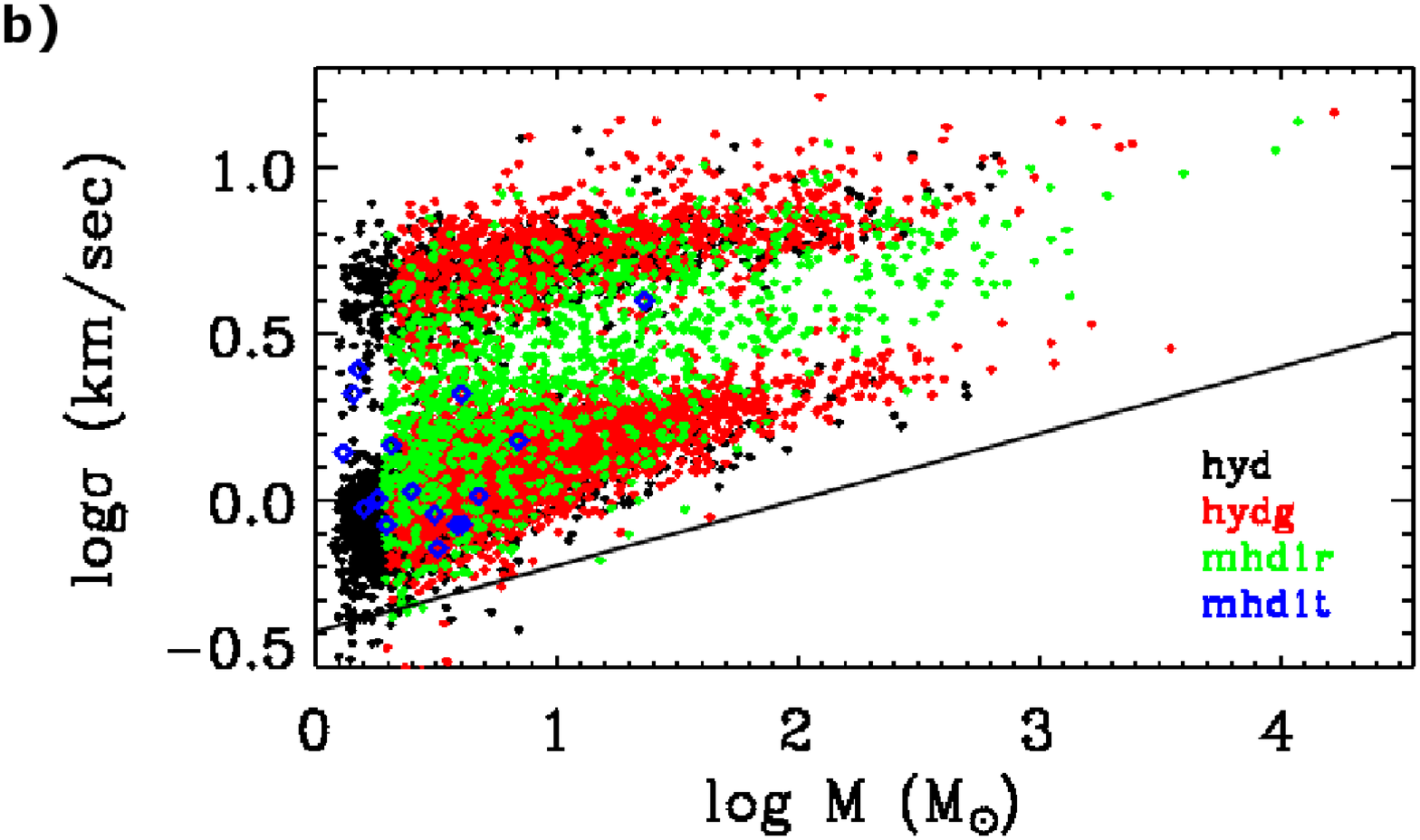}
    \includegraphics[width=0.49\linewidth]{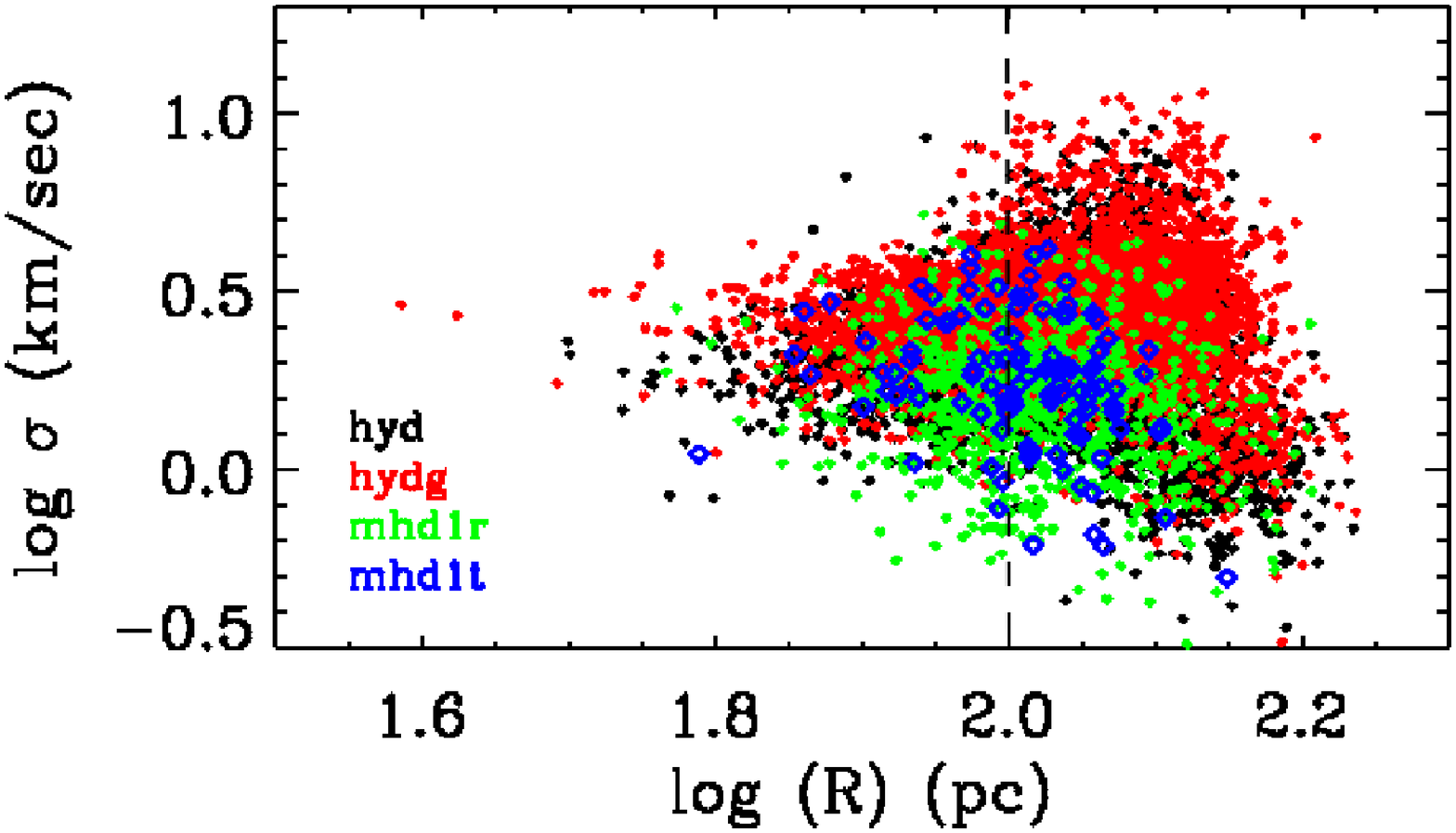}
    \includegraphics[width=0.49\linewidth]{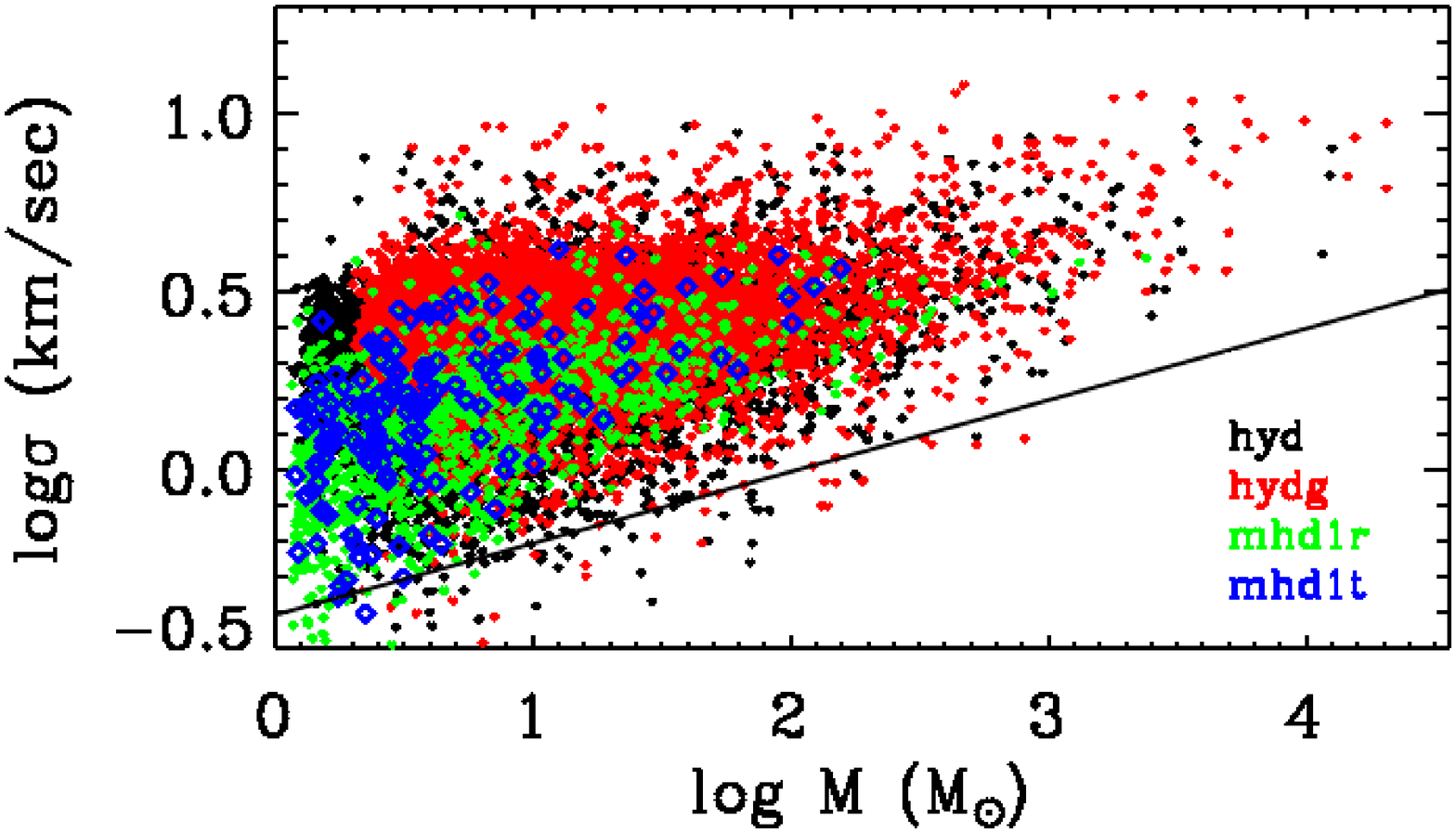}

    \caption{a) Three-dimensional velocity dispersion of the dense structures as function of the distance
from the closest superbubble, shown before the collision and at the end of the simulations.  The dashed lines show the mean radius of the bubbles in each run.  
The bottom plot shows snapshots after the collision, so there is only one dashed line, which indicates the middle of the box.  
b) Three-dimensional velocity dispersion of the dense structures as a function of their mass.  The solid line shows the Larson relation $\sigma~\text{(km/sec)}=0.4~M~(M_\odot)^{0.2}$ \citep{Larson_1981}.  
In both panels the top plots show times before the collision (t=2~Myrs for runs \textit{hyd}, 2.2 Myrs for runs \textit{hydg}, \textit{mhd1r,} and \textit{mhd1t}), and the bottom plots show times after the collision (t=7.6~Myrs for run \textit{mhd1t}, t=6~Myrs for runs \textit{hyd}, \textit{hydg,} and \textit{mhd1r}). Runs \textit{hyd} and \textit{hydg} are shown at slightly different times to avoid overlapping of the points.}
     \label{pos_vel_dis}
\end{figure*}
\begin{figure*}[th!]
    \includegraphics[width=0.49\linewidth]{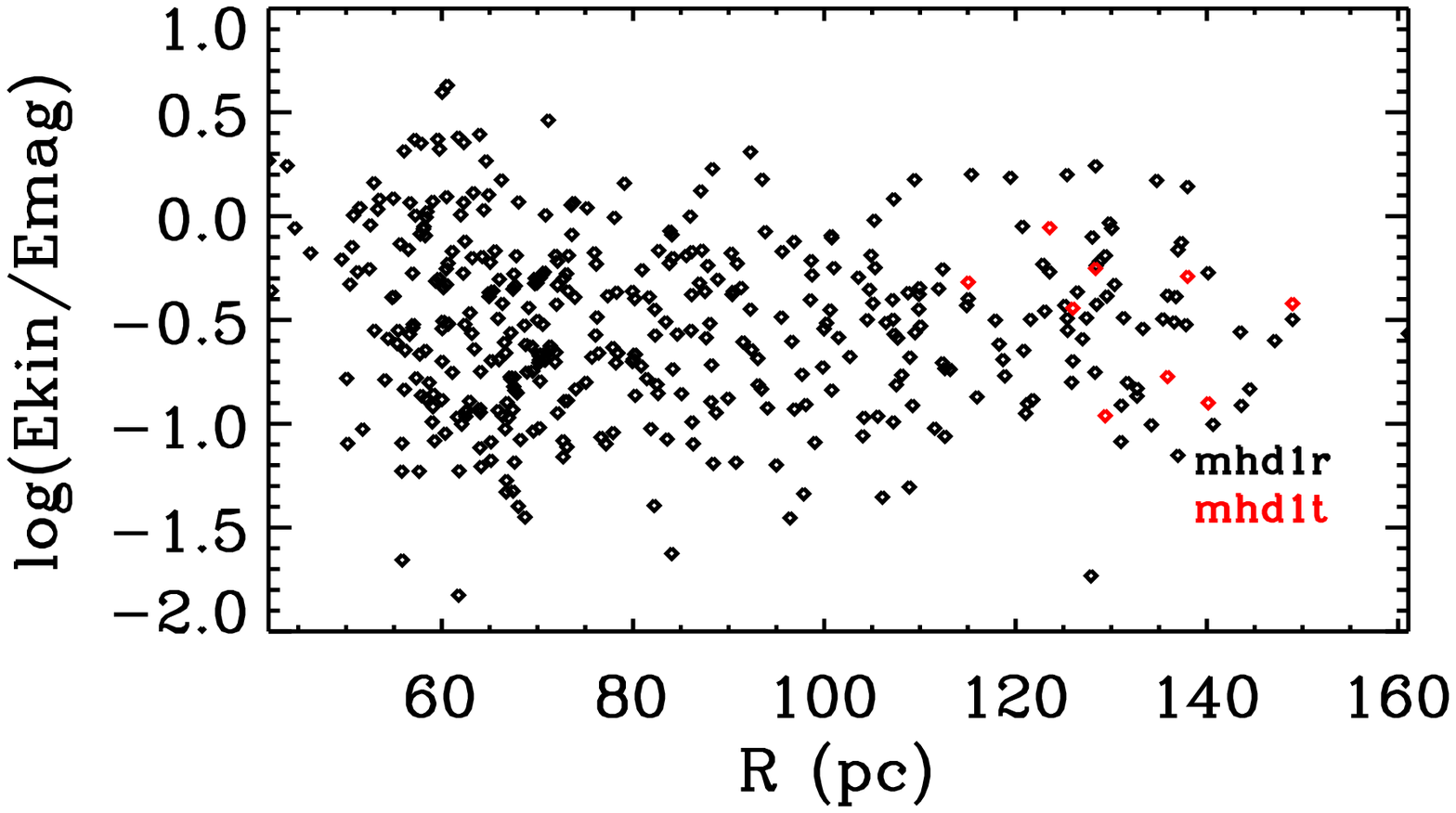}
    \includegraphics[width=0.49\linewidth]{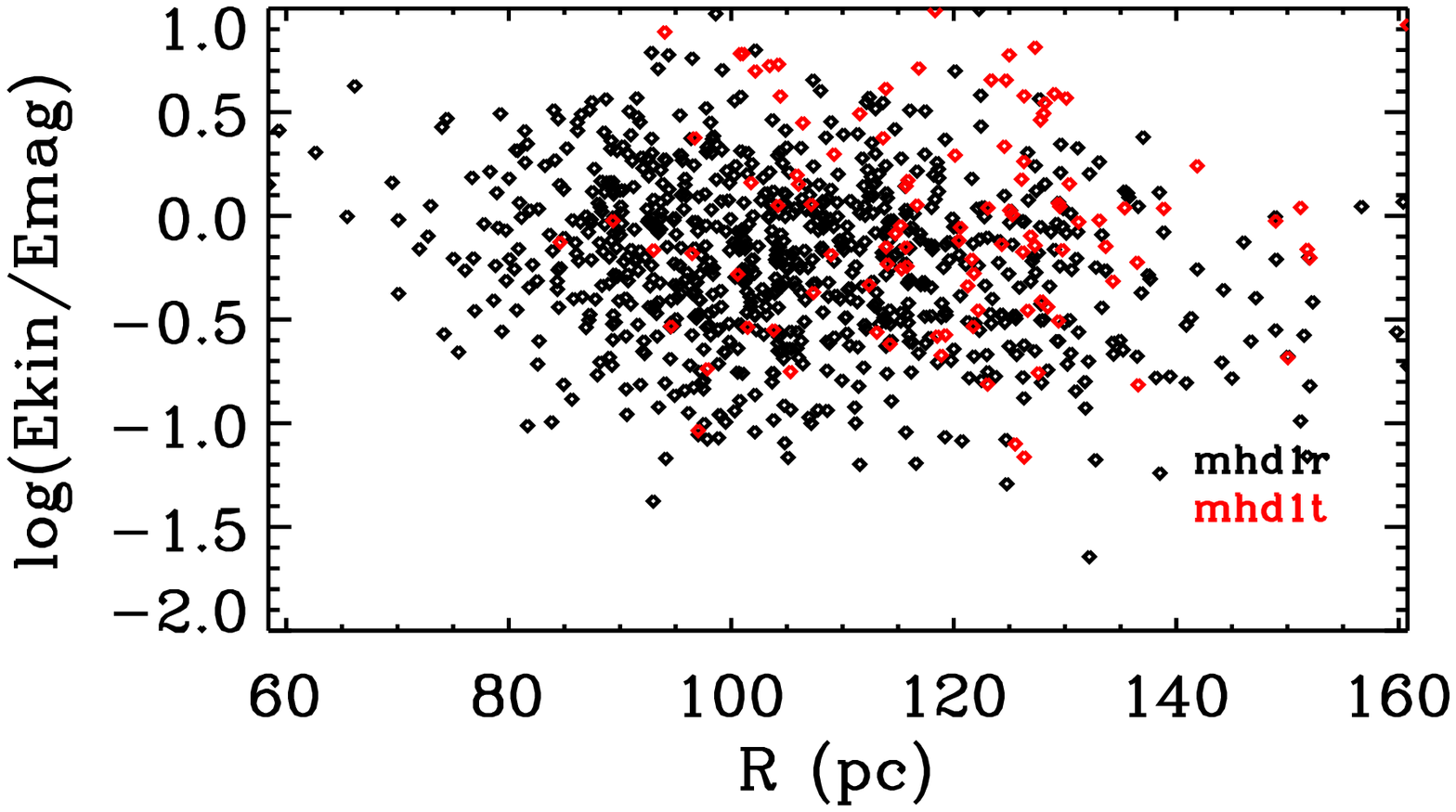}
    \includegraphics[width=0.49\linewidth]{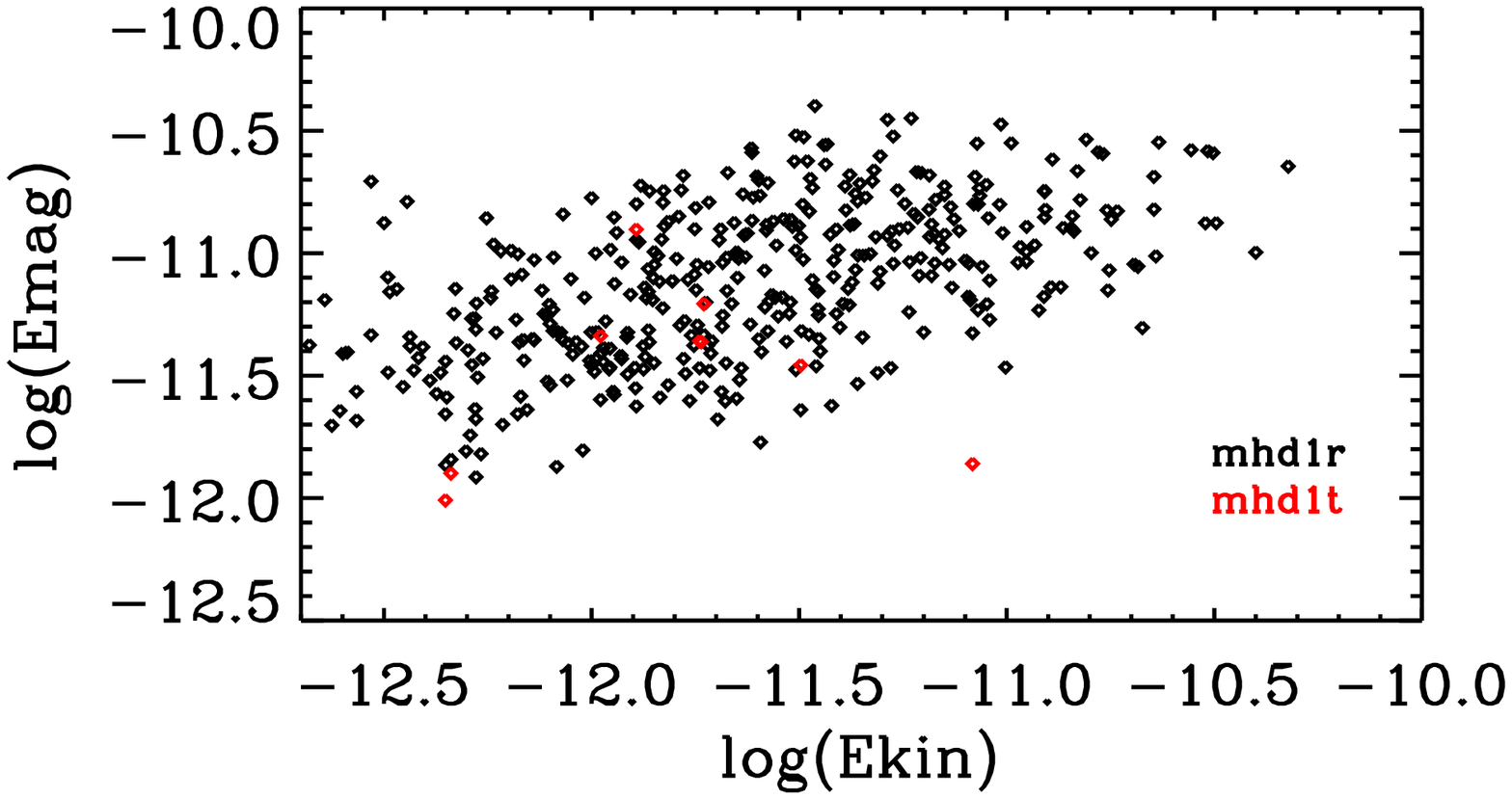}
    \includegraphics[width=0.49\linewidth]{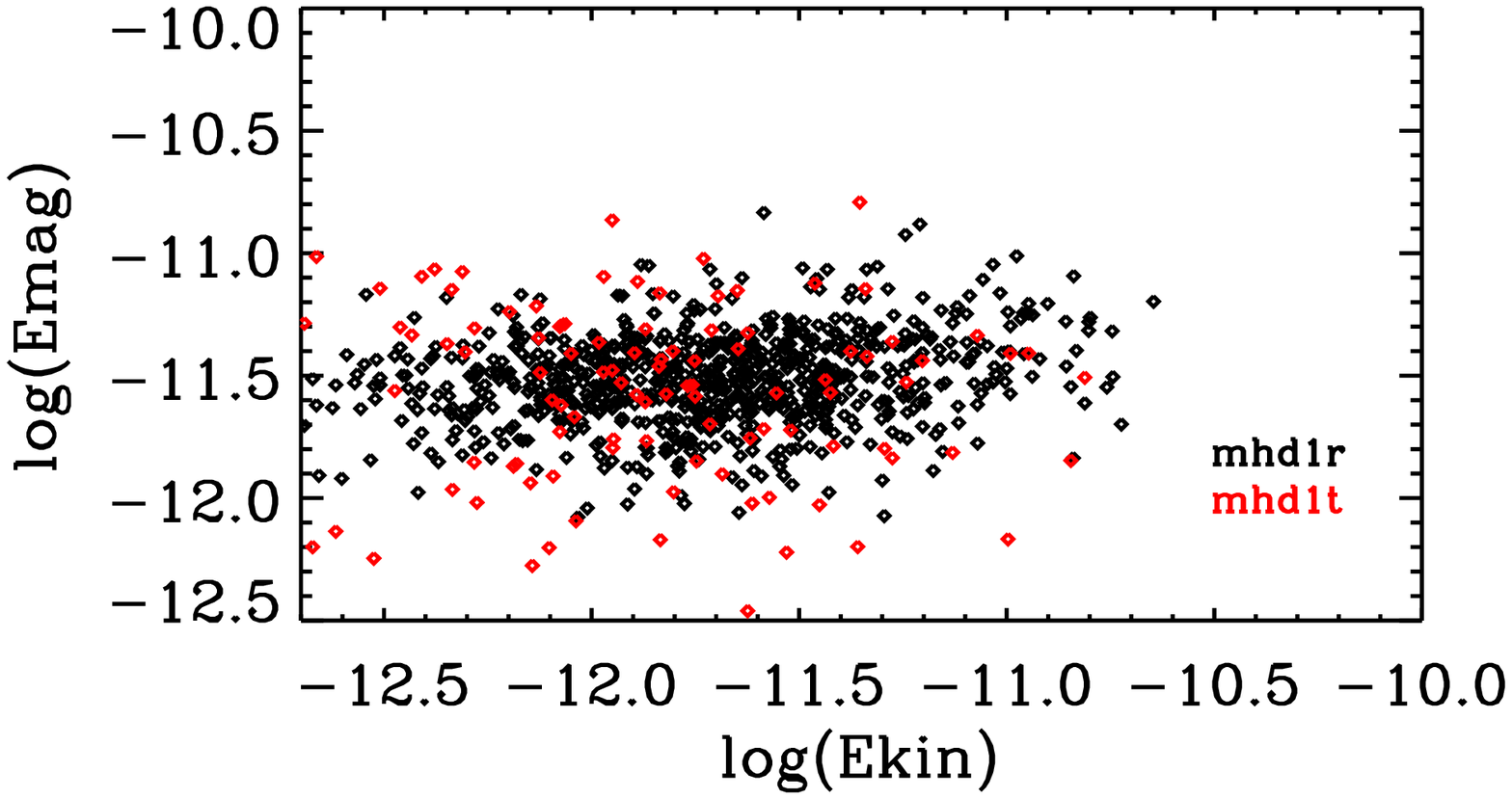}
    \caption{Top panel: Logarithm of the kinetic to magnetic energy ratio for cold clumps, which are defined as "friends-of-friends" groups
above a density threshold of 50 cm$^{-3}$, as a function of distance from the closest superbubble.  Bottom panel: Magnetic vs. kinetic energy 
of the clumps at the same times.  The black symbols correspond to run \textit{mhd1r}, the red symbols to run \textit{mhd1t}.
The snapshots are taken at times t=2~Myrs (left) and t=6~Myrs (right).}
     \label{ener_ratio}
\end{figure*}
%
\begin{figure*}[th!]
    \includegraphics[width=0.49\linewidth]{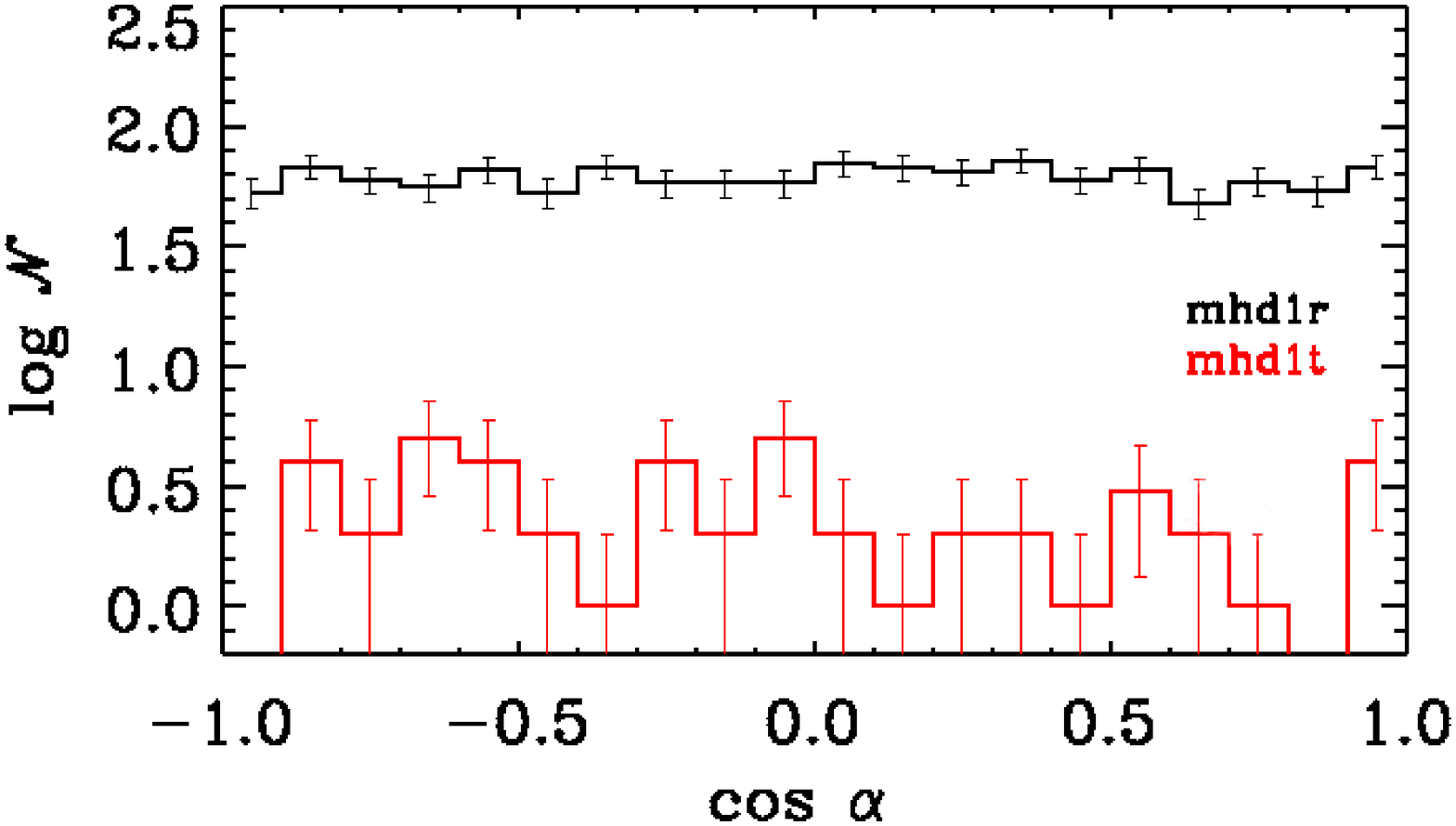}
    \includegraphics[width=0.49\linewidth]{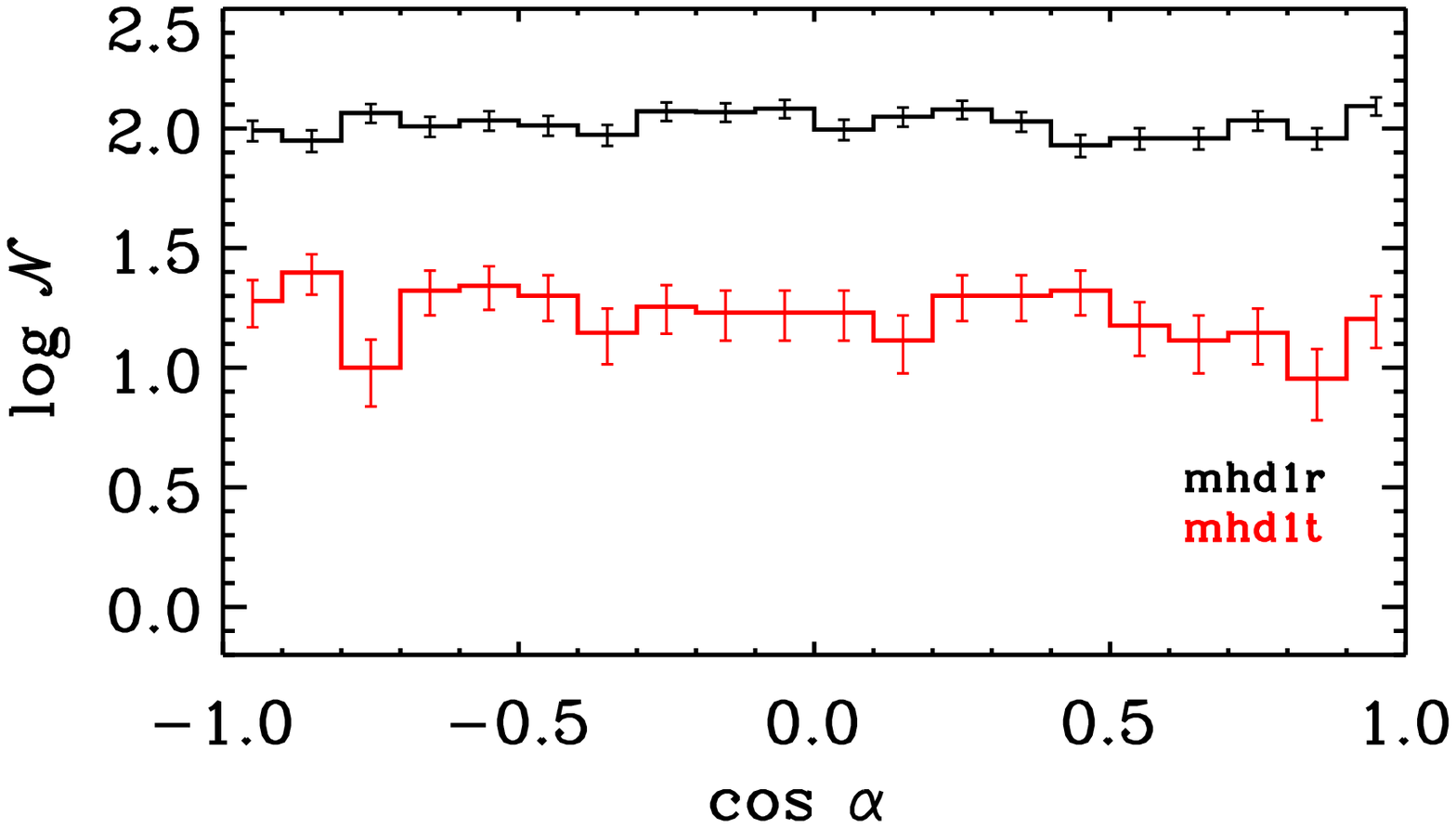}
    \caption{Distribution of the cosine of the angle between the main axis of the filament in three dimensions and the mean local magnetic field, 
at times t=2~Myrs (left) and t=6~Myrs (right) for runs \textit{mhd1r} (black) and \textit{mhd1t} (red).  Filaments are defined as described in the text and in the figures above.}
     \label{angle_dis}
\end{figure*}

%
\subsection{Evolution in a magnetized medium}

Four magnetized runs were performed for this study: \textit{mhd1r}, \textit{mhd1t}, \textit{single1l,} and \textit{single10l} (see Table \ref{tab_models}).  
Runs \textit{single1l} and \textit{single10l} are single bubble simulations and are presented in the Appendix to illustrate the effect
of increasing the magnetic field strength in an environment of almost uniform density.

Before the turbulence or feedback are introduced, 
the MHD runs all start with a uniform density of 1 cm$^{-3}$ and a uniform magnetic field in either the y or z direction.
As in the hydrodynamic runs, we drive turbulence for a few crossing times to reach statistical equilibrium before the feedback regions are included.
The collision in both runs occurs along the z-axis, so these two runs correspond to a magnetic field either perpendicular (run \textit{mhd1t})
or parallel (run \textit{mhd1r}) to the collision axis.
The magnitude of the initial, ordered magnetic field is 1.5 $\mu$G in run \textit{single10l}, and 5 $\mu$G in runs \textit{mhd1r}, \textit{mhd1t}, and \textit{single1l}.
The magnetic field thus develops a turbulent component, but part of the ordered mean remains from the initial configuration.

The middle and bottom panels of Fig. \ref{volume_rendering_all} show snapshots from the evolution of runs \textit{mhd1r} and \textit{mhd1t}, 
in the logarithm of the total gas volume density.  
Both snapshots of each simulation are taken before the shells collide and, as in the hydrodynamic cases, certain density ranges have been rendered transparent for illustration purposes.
The arrows show the local magnetic field, with sizes normalized to the highest value in the box.

Overall, we see that in both the magnetized cases the shells acquire a very different morphology with respect to the hydrodynamic situation.
The anisotropy introduced by the magnetic field affects the shape of the bubbles, but also affects the formation of dense gas, which appears 
organized into long filaments.  However, run \textit{mhd1r} is much more similar to the hydrodynamical run \textit{hyd}, since not only the shocks but also the environment becomes thermally unstable
under the pressure of the shocks.  Run \textit{mhd1t} forms significantly less dense gas because the configuration of the magnetic field allows material to escape from the 
upper and lower boundaries when compressed by the shocks.

The middle and bottom panels of Fig. \ref{shells_col_den_all} contain column density contours of runs \textit{mhd1r} and \textit{mhd1t}, respectively.
The snapshots shown are at about 2 Myr, before the shocks collide, and at the end of each simulation, 6 Myrs for \textit{mhd1r} and 7.6 Myrs for \textit{mhd1t}. 
The arrows show the mass-weighted projected magnetic field 
on the same planes.  Both the size and color of the arrows indicate the strength of the projected field in $\mu$G.
It is instructive to observe the relative orientation of the projected field with respect to the dense filamentary structures that appear in column density because this 
is a possible observable feature of such a system (for example, the results of \citealt{Clark_2014} for HI filaments, \citealt{Gaczkowski_2015} for the Lupus molecular cloud,
or statistically, the dust polarization results from \citealt{Planck_2016_2} and \citealt{Planck_2016}).  
When the magnetic field is perpendicular to the shell collision axis (run \textit{mhd1t}), there is a noticeable change both in strength and in orientation of the
projected field on the shock surfaces, even at early times.  However, when the magnetic field is oriented along the shell collision axis (run \textit{mhd1r}), this change is at the level of
the fluctuations due to the background turbulence.
At late times, in both models, due to the strong compression, the projected magnetic field at the center of the box is oriented mostly along the collision surface and its strength is significantly enhanced with respect to the field in the diffuse hot medium.

Next we look at the evolution of the shell radius with time in the different simulations.  
The analytical, self-similar description of the expansion of an adiabatic, spherical wind in a homogeneous environment gives a power-law  
dependence of the radius on time, $R(t)\propto t^{3/5}$ \citep{Weaver_1977}. To compare to this solution, in Fig. \ref{shells_radius_time} we plotted 
the logarithm of the shock radius versus the logarithm of time for runs \textit{hyd}, \textit{mhd1t}, and \textit{mhd1r}, and fitted the data points with a power law.
The slopes of the relation in runs \textit{hyd} and \textit{mhd1r} are almost exactly equal to ${3/5}$.
In contrast, the presence of a magnetic field 
perpendicular to the shell collision axis (\textit{mhd1r}) flattens the shock expansion law to about $0.4$.
A test simulation similar to \textit{mhd1t}, but with a mean magnetic field of 1.5$\mu$G, gave an expansion law with a slope of 0.45.
There are two reasons for this: first, the magnetic field provides an additional pressure in the postshock region 
and, second, because of the large anisotropies in the initial density of the environment a lot of the wind material can escape from the box through the boundaries.

Fig. \ref{shells_velocity_time} shows the radial velocity of the shells as a function of time in the different simulations.  
The radial velocities are calculated by taking the derivative of the curves in Fig. \ref{shells_radius_time}. 
In general, the expansion speeds are consistent with observed shells in the Milky Way \citep{Heiles_1979,Heiles_1984,McClure_2003} and in the Large Magellanic Cloud (LMC) \citep{Kim_99}, which typically range from 10 to 30 km/sec.  We see that runs \textit{mhd1r} and \textit{mhd1t} reach the sonic velocity of the WNM (on the order of 10~km/sec) much earlier than run \textit{hyd}, meaning that, when they eventually meet, their encounter is not supersonic like in \textit{hyd}.  This has significant consequences, for two reasons.  First, because any instabilities triggered by the collision, such as the nonlinear thin shell instability, are expected to evolve much more slowly at low Mach numbers \citep{Heitsch_2006} and second, because a less violent collision is not as efficient at destroying the pre-existing clumps by ram pressure stripping.

The density-pressure mass histograms in the center and right panels of Fig. \ref{TI}, which correspond to runs \textit{mhd1r} and \textit{mhd1t} respectively,
show a large fraction of the gas in the thermally unstable regime for both MHD runs.  
The same mechanism is at work here as in the hydrodynamical situation.The gas condenses, moves away from the WNM equilibrium, and ends up
in a region where cooling dominates, namely the thermally unstable regime.  Here, this effect is amplified with respect to run \textit{hyd} owing to the additional pressure from the magnetic field.

In general, the MHD runs produce less cold and dense gas compared to the unmagnetized models.
We can get a clearer estimate of this by defining three temperature regimes:
cold for temperatures below 100~K, warm for temperatures between 100 and 10000~K, and hot for temperatures above 10000~K,
and plotting the mass in each of these regimes. This is shown in the left panel of Fig. \ref{gas_phases}. In this figure
the diamonds correspond to run \textit{hyd}, the crosses to run \textit{mhd1t}, and the triangles to run \textit{mhd1r}.
Blue lines correspond to the cold gas, green lines to the warm, thermally unstable gas, and red lines to the hot wind material.

There is a clear conversion of warm to cold gas in run \textit{hyd}.  After about 3~Myrs and for the rest of this simulation
the cold gas accounts for most of the mass. 
None of the MHD runs ever reach as high a cold gas fraction of dense gas  because the magnetic field stabilizes the gas against the TI.
This is even more evident in the plot of the middle panel, which shows the total mass in the box as a function of time and the absolute values of the cold gas mass.  
After 6 Myrs, run \textit{hyd} has produced more than $10^5$ M$_{\odot}$ of dense gas, compared to only $10^{3.5}$ M$_{\odot}$ in run \textit{mhd1t}.

The right panel of Fig. \ref{gas_phases} shows the fraction of kinetic energy carried by each gas phase.  
Interestingly, it appears that, not only the presence of a magnetic field, 
but also the different magnetic field configurations cause a different distribution of mass and kinetic energy.
In the hydrodynamic case, most of the total kinetic energy is in the cold phase, 
while in runs \textit{mhd1r} and \textit{mhd1t}, the bulk of the kinetic energy lies in the warm, thermally unstable gas.
In run \textit{mhd1t}, the cold phase barely carries any kinetic energy, while the hot phase achieves the highest
kinetic energy portion among the three simulations.

To better understand the dynamics, in Fig. \ref{ener_time} we plot the total kinetic and magnetic energies in the three runs \textit{hyd}, \textit{mhd1r}, and \textit{mhd1t} as a function of time. 
While the total amount of kinetic energy in the box remains roughly constant in both MHD runs, in \textit{hyd} it increases to about five times the initial value at 4 Myrs (about the time of the shell collision) and only
starts decreasing toward the end of the simulation. Both MHD runs show an important 
increase in magnetic energy early on, but that is not enough to account for the constancy in the total kinetic energy. 
This implies that in the hydrodynamical situation the kinetic energy inserted by the feedback regions remains in the box, while in the MHD runs 
the kinetic energy escapes from the boundaries early on because of the elongation of the superbubbles.


\section{Dense structure: Clumps and filaments}
\label{sec:filaments}


A simple visual inspection of Fig. \ref{volume_rendering_all}
shows that the dense structures formed in the magnetized runs are more filamentary than in the 
hydrodynamic case, where we observe almost spherical clumps.  
In this section we examine and compare the properties of these dense formations in more detail.

Clumps and filaments are identified by linking cells with densities above a certain threshold 
(here 50~cm$^{-3}$) with a friends-of-friends algorithm.  The threshold is chosen to correspond to a value slightly higher than the minimum of the cooling-heating equilibrium curve that
is responsible for the formation of a two-phase medium.

The mass distributions of the clumps in runs \textit{hyd}, \textit{hydg}, \textit{mhd1r,} and \textit{mhd1t} are shown
in Fig. \ref{mass_histo} for two stages of the simulation: at 2~Myrs, before the bubbles collide, 
and at the end of each simulation.
At both evolutionary stages there is a very small difference between runs \textit{hyd} and \textit{hydg} (upper panel of the figure),
since the self-gravity in run \textit{hydg} creates slightly more massive structures.  However, this difference is
within the statistical errors. A more marked difference is between hydrodynamics and MHD, 
since runs \textit{mhd1r} and \textit{mhd1t} contain
structures of much lower masses in comparison to the situation of pure hydrodynamics.  
Of the two magnetized runs, \textit{mhd1t} forms overall fewer dense filaments.   

Fig. \ref{axis_ratios} shows the axis ratio distribution of the identified dense structures. Here, 
the principal axes of each structure are calculated as the eigenvectors of its inertia matrix and the axis ratios are the ratios of the corresponding eigenvalues. 
The top plot shows the distributions before the collision, and the bottom one after it. 
Unfortunately, the statistics for run \textit{mhd1t} are very poor, so that little can be concluded about the general trend of the clump shape.
The remaining tendencies among runs runs \textit{hyd}, \textit{hydg}, and \textit{mhd1r} are weak.
At early and late times alike, all three runs produce a power-law distribution of axis ratios, with a gradual cutoff toward
low ratio values, as clumps become more poorly resolved.  The few resolved dense clumps of run \textit{mhd1t} show 
a relative shift between the distributions of second longest to longest, and of shortest to second longest axes, which could be interpreted
as an excess of filamentary structures.

As we saw in the previous section, the kinetic energy is distributed differently in the different simulations.  
It is therefore very interesting to compare the internal velocity dispersion of the cold structures between different simulations.
In Fig. \ref{pos_vel_dis} we plot the three-dimensional velocity dispersion within the cold structures
as a function of their distance from the closest superbubble and as a function of their mass.  
At early times (t=2~Myrs, top), the two hydro runs seem to produce two clump populations: one at low ($\sigma<2$ km/sec) and one at high velocity dispersion ($\sigma\simeq 4-10$ km/sec). 
The population at high velocity dispersion is located closer to the mean radius of the superbubbles, indicated with dashed lines. 
This is much less pronounced in the magnetized runs. At the end of the simulations this separation of populations disappears.

We interpret the appearance of clumps with high velocity dispersion close to the supershells as evidence of the Vishniac instability, which is very active in the hydro simulations, but largely suppressed in the MHD simulations.  The clumps that form on the shocks due to the combination of the Vishniac and the TI inherit the velocity of the shock, which triggers strong turbulence in their interior.   
In contrast, clumps that form far away from the shocks due to the overall increase of  pressure in the volume only carry the turbulent velocities of warm gas.

We also notice that there is a clear correlation between mass and velocity dispersion.  However,  
all the dense structures are always well above the Larson relation $\sigma~\text{(km/sec)}=0.4~M~(M_\odot)^{0.2}$ \citep{Larson_1981}, which is indicated in the plots with a solid black line.  
This means that these clumps are not representative of molecular clouds in the ISM, at least not in terms of their velocity dispersion.  
It is likely that they lose kinetic energy gradually as the shock compression subsides, or as they cool further 
and condense into the molecular phase, but we do not model these processes in our simulations.

Given the marked difference in the behavior of the MHD simulations, it is useful to see how the kinetic energy compares to the magnetic energy in these filaments.
In Fig. \ref{ener_ratio} we show the kinetic versus the magnetic energy of each structure in runs \textit{mhd1t} and \textit{mhd1r}, 
and the ratio of the two energies as a function of distance from the closest superbubble; at 2~Myrs (left-hand side),  the average shock radius for both runs is 45 pc.
The dense structures in run \textit{mhd1r} are found all over the computational volume, while those in \textit{mhd1t} are located further away from the shocks.
Also, the energy ratios increase slightly with time.  As we can see from the E$_{mag}$-E$_{kin}$ plots, this does not appear to be from an increase in their kinetic energy, but is rather from a decrease in their magnetic energy.  
How or why these structures lose their magnetic field is not easy to answer.
One possibility is that structures merge, or get destroyed and reform, so their magnetic field is redistributed.
Another possibility is that numerical diffusion or reconnection are responsible for this redistribution.
However, with these simulations alone we cannot distinguish between these different scenarios.  We would need a parameter study of diffusion values, and a much higher resolution, both of which are computationally very costly in this setup.

Finally, we look at the distribution of the cosine of the angle between the main filament axis and the local magnetic field, $<\text{cos}\alpha>$, in three dimensions (Fig. \ref{angle_dis}).
This quantity is calculated as the mean of 
\begin{equation}
\text{cos}\alpha = \frac{{\bf{B_l}}\cdot{\bf{r_1}}}{|{\bf{B_l}}|~|{\bf{r_1}}|}
\label{cos}
,\end{equation}
where ${\bf{B_l}}$ is the magnetic field in a cell belonging to the filament, and ${\bf{r_1}}$ is the local direction of the filament.
As we explained, the main axis of the structures is defined as the eigenvector of the inertia matrix that corresponds to the largest eigenvalue.
In order to calculate quantities along its length, the structure is sliced along that main axis 
and local centers of mass are found for each slice. In Eq, (\ref{cos}), ${\bf{r_1}}$ is essentially the vector connecting two adjacent barycenters.
The left panel shows histograms of t=2~Myrs, and the right panel shows histograms of t=6~Myrs.  The main filament axis is determined by the largest moment of inertia of each structure,
which is the largest real eigenvalue of its inertia matrix.
According to this plot, there is no visible preference for alignment or misalignment of the structures with the local magnetic field in either run and no trend with 
time. 


\section{Summary and discussion}
\label{discussion}

We presented simulations of the expansion and collision of two superbubbles in a diffuse turbulent ISM with which
we examined the role of magnetic fields in the expansion and collision of supershells.

Little, if any, differences are found between hydrodynamical simulations with and without self-gravity.
This is consistent both with the expectations from the linear analysis \citep{Vishniac_83},  
and with previous numerical simulations, which only included gravity \citep{Palous_2003}.
According to the linear theory, the fact that the condensations happen on a shell does not accelerate the gravitational instability.
In other words, scales that were gravitationally stable remain so even under the influence of the shocks.
This is evidenced in the simulations by the fact that only the very highest densities seem to be affected by gravity.
These high densities occur only after the shock has fragmented dynamically.

In the magnetized case, superbubbles expand faster along the magnetic 
field lines than perpendicular to them
owing to the magnetic pressure confining the shells.
This additional pressure also stabilizes the shocks against the fluid instabilities
so clearly at work in the hydrodynamic case.  This translates into an overall decrease of the dense gas fraction with respect to the purely hydrodynamic situations.

Upon visual inspection, the magnetized simulations appear to produce more elongated, filamentary clumps than the hydrodynamical simulations.
This is in agreement with the results of \citet{Hennebelle_2013}, who showed that MHD turbulence produces more filamentary structures
than hydrodynamic turbulence.
In our statistical analysis, there are some indications of an excess of elongated structures in the 
magnetized runs, but the statistics is not enough to suggest a systematic strong difference in morphology.

A closer examination of the dense structures reveals clear differences between hydrodynamical and MHD
situations.  
At early times, the cold filaments in the MHD runs have, on average, lower internal velocity dispersions compared to the hydrodynamical models.
Interestingly, the cold clumps in the hydrodynamical simulations continue growing in mass until late times, 
while in the MHD simulations they seem to stay within the same mass range.

In the MHD simulations we observe that the kinetic to magnetic energy ratio of the clumps
increases with time.  This was also reported by \cite{Inoue_2012} in simulations of colliding HI flows, where they observed
an increase of the clump masses with time and, therefore, an increase in their velocity dispersion according to the Larson relations. 
However, it is unlikely that this is the reason behind the increase of the kinetic to magnetic energy ratio here.
As we saw, the clumps in our MHD simulations do not grow significantly in mass and they do not show any notable increase in kinetic energy.
Instead, they gradually lose their magnetic field during the shock collision.  It is unclear whether this happens because of diffusion 
or reconnection.

The formation of cold filaments in the magnetized runs is reminiscent of the gravitational fragmentation of a thin layer 
threaded by a magnetic field, as predicted by \citet{Nagai_1998}.
They found that for high enough magnetic field strengths, the tension perpendicular to
the mean field direction resists gravity and prevents collapse along the field lines, thus forming elongated dense structures. 
In our simulations of magnetized shells there is no gravity, yet 
there is an acceleration on the surface of the shell from the expansion; this acceleration is balanced by the magnetic tension,
causing the formation of these very long filaments.

The hydrodynamic simulations with and without self-gravity are the same simulations \citet{Dawson_15} used 
for a comparison with a molecular cloud compressed by two supershells.
The conclusion of that work does not change with the inclusion of magnetic fields.  In fact, the total amount 
of dense gas is up to two orders of magnitude less in the magnetized compared to the unmagnetized situation.

Based on these simulations, and without ignoring the limitations of the parameter space and the initial conditions,
it is interesting to attempt to understand the conditions in observed molecular clouds that are located between superbubbles.
For example, \citet{Fujii_2014} find dense molecular clumps at early stages of star formation between the LMC~4 and LMC~5 supershells.
The expansion speeds of these superbubbles, which are 21 and 36 km/sec, respectively \citep{Kim_99}, are similar to what we get in our simulations, but 
their observed sizes are much larger (i.e., radii of about 290 and 550 pc, assuming a distance of 50~kpc to the LMC). This suggests that they 
either expanded in a lower density, lower magnetization medium than what we simulate, or that the stellar populations that created them 
produced more powerful feedback.  
Whatever the case, the existence of dense, star-forming gas between two high-Mach shells 
appears in some contradiction with our findings in this paper and in \citet{Dawson_15}.  
The same is true of galactic shell collisions. \citet{Gaczkowski_2015} argue that the star formation properties of the Lupus I cloud 
are consistent with it being squeezed by stellar feedback from the Upper Scorpius and Upper Centaurus-Lupus regions.  These shells have expansion velocities of
about 20-30 km/sec \citep{Heiles_1979}, which are supersonic with respect to the WNM of the Galaxy.
Instead, our simulations show that the amount of dense gas after 
the supershell collision is not higher than the sum of the dense gas formed by each individual shell.  
This effect is stronger with magnetization, in which case the shock compression and, consequently, also the dense gas formation, are largely suppressed.

However, this is not necessarily inconsistent with shock-triggered molecular cloud formation. 
\citet{Inutsuka_2015} argue, for example, that binary shock collisions are sometimes insufficient to form dense molecular
clouds and that multiple compressions may be needed for the phase transition to happen.  
This is also hinted in the results of \citet{Walch_15}, who, when simulating a galaxy with different star formation and feedback recipes, 
observe a delay in the formation of molecular gas in the presence of a magnetic field.
Although our work does not investigate shock collisions in a multi-phase medium or explore the parameter space to the extent necessary for making a general statement, 
it shows that shock collisions do not necessarily form enough dense gas to create a molecular cloud directly out of the WNM.

A very interesting discussion has opened lately, following the
observational finding that low-density molecular   filaments and diffuse HI 
filaments are mostly aligned with the projected magnetic field \citep{Soler_2013,Clark_2014, Planck_2015,Planck_2016}.
Could these features be filaments on the walls of superbubbles or other large-scale shocks?
Although we have not performed a detailed analysis in this work, our models show no preferred orientation of the magnetized filaments relative to
the local magnetic field.  
The reason for this difference is not clear.  One possibility is that the observed filaments are not formed on the surface of a shock, but by a different process, 
such as isotropic MHD turbulence.  Or, if they are indeed formed in shocks, these shocks in nature might be moving with a higher Alfven Mach numbers than in our simulations.
We postpone this investigation until we can perform a wider parameter study.


\section{Conclusions}
\label{conclusions}
We conclude that the structure and collision dynamics of supershells are both greatly affected by the presence of a magnetic field,
whether it is mostly oriented parallel or perpendicular to the axis of collision.
The additional pressure and tension provided by the magnetic field stabilizes the shells against the Vishniac instability and thus also hinders the formation of dense gas through TI. 
In addition, the gas that does become thermally unstable condenses preferentially in the form of filaments rather than 
quasi-spherical clumps.  

In the hydrodynamic models the densest material is clearly associated with the shell surface.  While in all the models the compression by the shells pushes the gas into the thermally unstable regime, in the magnetized models the multi-phase ISM appears similar to a generic ISM region
and does not show any direct association between dense gas and supershells.
This means that in the presence of a magnetic field the shell expansion has a much less obvious impact on the surrounding gas and would be easily missed in observational surveys.

Finally, we have shown that shell collisions are not very efficient in forming additional dense gas with respect to what the shell fragmentation
alone has created.  What is more, in the cases without magnetic field or with a magnetic field along the collision axis, the collision speed is supersonic, therefore
potentially high enough to destroy some of the previously existing structure.   

Although the parameter space of supershell collisions still needs to be explored, our results show that a simple collect-and-collapse scenario
or a collision-triggered molecular cloud formation scenario cannot always explain the presence of molecular clouds around or between supershells.

\emph{Acknowledgments}

We would like to thank Shu-ichiro Inutsuka and Juan Diego Soler Pulido for very useful discussions on this work.  
We are also grateful to the anonymous referee for constructive comments and suggestions that greatly improved this manuscript.
This research has received funding from the European Research Council under the European Community's Seventh Framework Programme (ERC Grant Agreement "ORISTARS", no. 291294 and FP7/2007-2013 Grant Agreement no. 306483). 

\bibliographystyle{apj}
\bibliography{bubbles}
\label{lastpage}


\appendix

\section{Evolution of a single bubble in a uniform magnetic field with low turbulence}
\label{app:one bubble}

\begin{figure}[!h]
   \includegraphics[width=\linewidth]{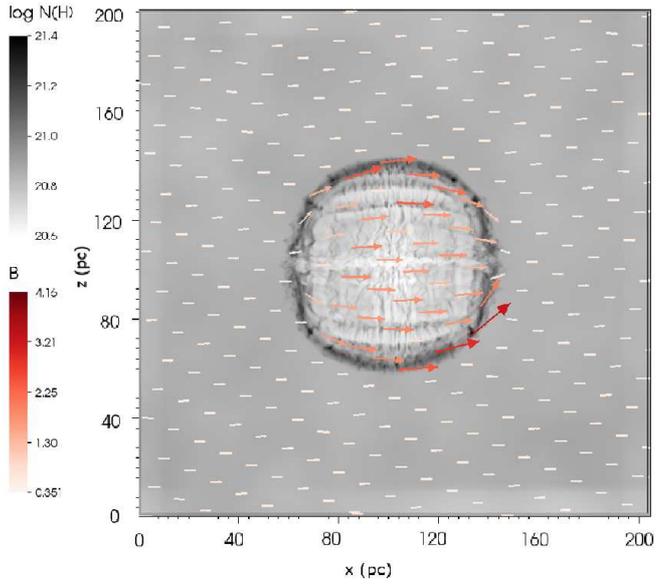}
   \includegraphics[width=\linewidth]{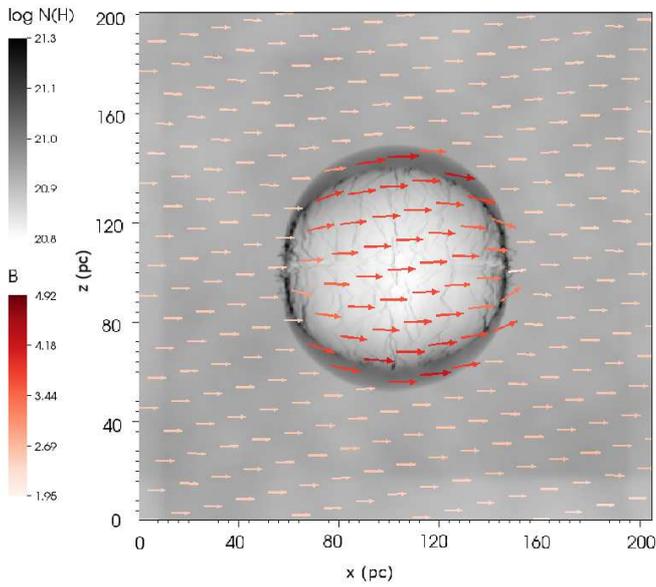}
   \caption{Expansion of a single bubble in a low magnetic field (top, plasma $\beta$=10) and  high magnetic field 
  (bottom, plasma $\beta$=1, analogous to runs \textit{mhd1r} and \textit{mhd1t}) with much lower levels of turbulence.  
The image in gray scale is the logarithm of the hydrogen column density in log~(cm$^{-2}$) and 
the arrows show the projected magnetic field, color coded according to its magnitude in $\mu$G.  
 The plots correspond to t=0.2~Myrs for both simulations.}
   \label{single_bubbles}
\end{figure}

\begin{figure}[!h]
   \includegraphics[width=\linewidth]{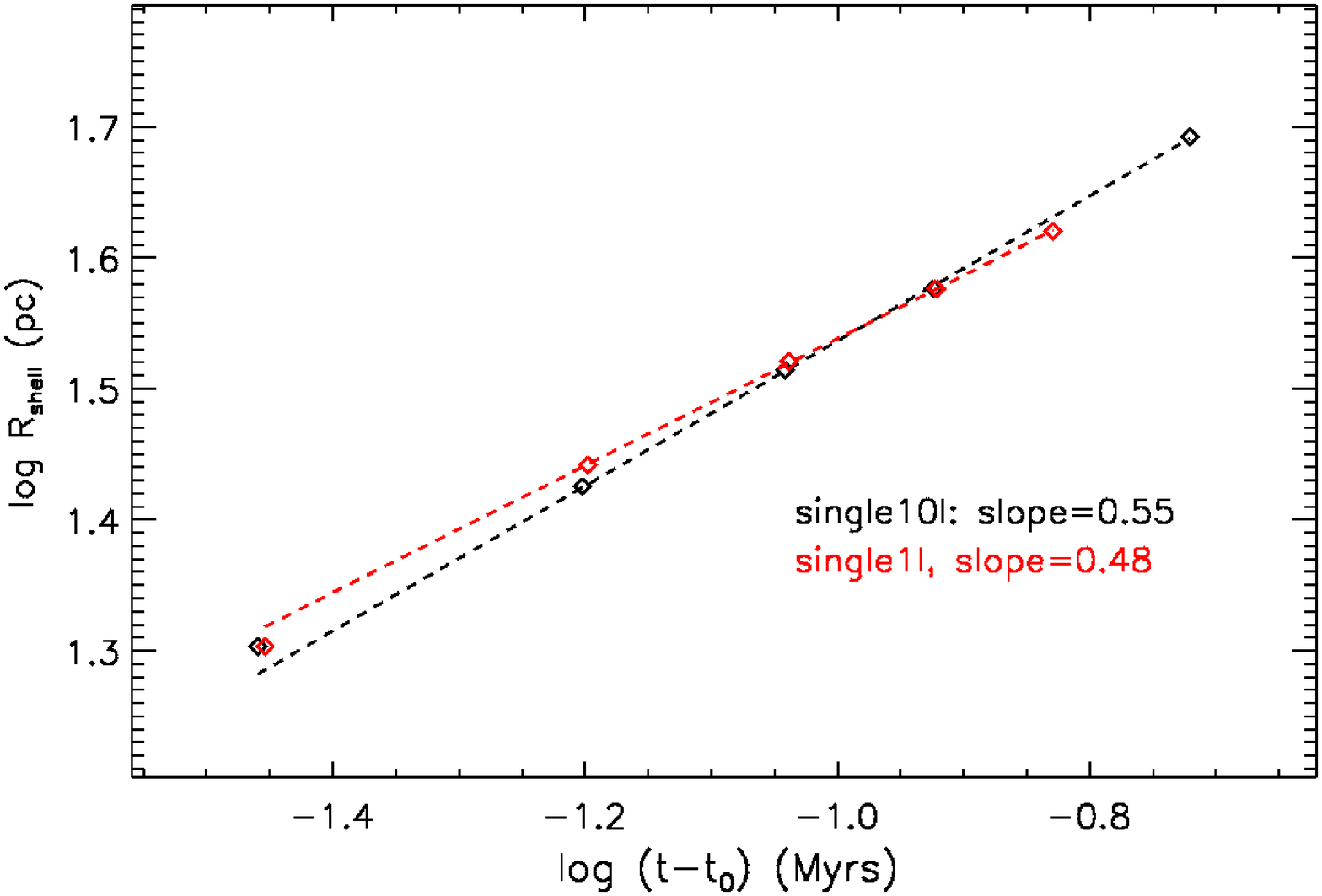}
   \caption{Logarithm of the average shell radius vs. logarithm of time for the single bubble models.
Time is plotted relative to the first snapshot available from the simulations, which is at about 0.1~Myrs}.
The dashed lines are linear fits with the slope mentioned on the plot.
   \label{law_single_bubbles}
\end{figure}

It is instructive to see what the expansion of a single bubble looks like in environments with different magnetizations when the effects of turbulence are ignored.  
To illustrate this we performed two simulations with a single superbubble,
which have a very similar setup to those presented in the main body of the paper.
The number of stars inside the superbubble and the mean temperature and density of the environment are the same, but the sonic rms Mach number is 10 times lower. 
In this sense, the turbulence is only there to provide the low-amplitude perturbations for triggering the shell instabilities. 

Fig. \ref{single_bubbles} illustrates the results.  
When the magnetic field is stronger, the additional magnetic pressure causes the shell to thicken perpendicular to the mean
field direction.  The shell also fragments less compared to the simulation with lower magnetization. 

The expansion law of the bubbles in this case is shown in Fig. \ref{law_single_bubbles}.

\end{document}